\documentclass[%
reprint,
floatfix]{revtex4-2}
\usepackage[inline]{enumitem}
\usepackage{dcolumn}
\usepackage{comment} 
\usepackage{xcolor}
\usepackage{graphicx}
\usepackage{subfigure}
\usepackage{amsmath}
\usepackage{physics}
\usepackage{amssymb}
\usepackage{bm}
\usepackage{hyperref}
\usepackage{blindtext}
\hypersetup{
    colorlinks=true,
    linkcolor=blue}
    
\begin{document}
\title{U(1) quantum spin liquids in dipolar-octupolar pyrochlore magnets: a fermionic parton approach}


\author{Krushna Chandra Sahu}
\email{krushnasahu@students.iisertirupati.ac.in}
\affiliation{Department of Physics, Indian Institute of Science Education and Research (IISER) Tirupati, Tirupati - 517507, Andhra Pradesh, India}
\author{Sambuddha Sanyal}\email{sambuddha.sanyal@iisertirupati.ac.in}
\affiliation{Department of Physics, Indian Institute of Science Education and Research (IISER)
Tirupati, Tirupati - 517507, Andhra Pradesh, India}
\date{\today}


\begin{abstract}
We study the uniform $U(1)$ quantum spin liquid (QSL) with low-energy fermionic quasiparticles for pyrochlore magnets with dipolar-octupolar symmetry, employing a fermionic parton mean field theory approach. Self-consistent calculations stabilize 12 fully symmetric uniform $U(1)$ QSLs; of which four mean-field states are ”monopole-flux” states. Several of these mean-field states show a linear temperature dependence of specific heat at low temperatures, the other phases show a power law temperature dependence of specific heat $C \sim T^\alpha$, where $\alpha $ is close to 1. We further compute the dynamic spin structure factors and
discuss the possible signature of these fermionic spinons
in neutron-scattering experiments on DO magnetic systems. Our results provide a possible way to understand the metallic specific heat response in $Nd_2 Sc NbO_7$.
\end{abstract}
\date{\today}
\maketitle

\maketitle


\section{Introduction}
Fractionalization in quantum spin liquids (QSLs) is one of the outstanding quests of modern condensed matter physics. Fractionalization is well established in one-dimensional spin systems\cite{1dchain_spinonexpt,1dchain_spinonexpt_1} and two-dimensional semiconductors in the presence of a strong magnetic field in the form of fractional quantum hall effect\cite{fqhe_nobel}. On the other hand, Quantum Spin Liquids(QSLs) are believed to be a rich playground of fractionalization physics due to many possible material realizations.  QSLs are ground states of magnetic systems that are long-range entangled and preserve the symmetries of the original Hamiltonian. Such symmetry-preserved (or enhanced) ground states can't be described in the paradigm of Landau's symmetry-breaking phases as they don't have any local order parameters; these states are described as "quantum-ordered" states. A distinguished feature of quantum-ordered states is the presence of fractionalized, non-local excitations; out of all the features of such states, this is most important due to its potential experimental relevance. 

Theoretically, the search for QSL ground state starts at magnetic Mott insulators where the effective degrees of freedom are the magnetic moments; thus, a spin model describes low energy physics. In a generic spin model, the ground state could be either a long-range ordered, symmetry-breaking state with gapless Goldstone modes or a symmetry-preserving state. The symmetry-preserving ground state can be either unique and paramagnetic with gapped excitations or degenerate, long-range entangled with gapped or gapless excitations. Lieb-Schultz-Mattis theorems\cite{LSM} (and extensions/generalizations\cite{oshikawa_lsm,hastings_lsm,wen_lsm,Parameswaran2013} thereof) rules out the possibility of trivial (short range entangled) paramagnet in certain cases but for the rest all the possibilities can occur mutually exclusively, in recent works a possible co-existence of long-range order and long-range entangled states were also discussed\cite{sambuddha_pyrochlore,Chern2019}. Typically, a symmetry-broken state can be ruled out in a model with highly frustrated interactions; in such cases, one expects the ground state to be a superposition of a macroscopic number of states in a highly degenerate manifold. The presence of gapped/gapless excitations and topological characteristics of the ground state can be confirmed by computing the entanglement properties with a number of theoretical tools such as DMRG\cite{davidhuesDMRG,davidhuesDMRG2}, ED\cite{Lauchli}, but these approaches are limited in lower dimension. A powerful approach that can be used in 2D\cite{kagome-triangular-psg, Bieri-chiral} as well as in 3D\cite{Liu_fermion,Liu_boson} is the parton(or spinon) construction approach,  where one expresses each spin degrees of freedom in terms of two fermionic or bosonic degrees of freedoms as if they are parts of the spin degree of freedom, those partons are by nature non-local and fractional and requires an additional gauge field in the description for a faithful representation. Expressing one degree of freedom in terms of another is a convenient tool for solving strongly correlated problems. However, in some of these cases, the partons can be deconfined i.e., they can be the true quasi-particles of the system and can leave their impressions in experiments.  In a seminal work, X.G. Wen\cite{wen_original} developed a systematic approach to construct all the possible symmetry-preserving parton constructions, which can be solved with a mean field approximation to get traction on developing a variational ansatz for the possible fractionalised QSL ground state of a given frustrated spin model.
If deconfined partons are the true quasiparticles of the system, then one expects those mean-field solutions to be stable beyond mean-field corrections such as the gauge fluctuations. These parton mean-field solutions are often a useful "first responder" approach towards understanding the possible landscape of QSL phases in various models of frustrated magnets\cite{balents2010-review,Savary_2017,zhou-review,broholm-review, gaudet2019quantum,benton2015ground,buessen2018quantum,das2019nirh}. 

In this work we focus on the dipolar-octupolar (DO) pyrochlore system\cite{rau-review}. DO pyrochlores can be described by an XYZ spin interaction model with a pseudo spin 1/2 magnetic moment at the vertices of a pyrochlore lattice, where two components of the magnetic moment transform as a dipole and one transforms as an octupole. The rare-earth pyrochlore compounds are one of the most commonly found frustrated magnets in nature and are considered strong candidates to host QSLs in 3D.  In very recent times several rare-earth pyrochlore compounds have emerged where the magnetic moment-inducing atoms have a low-lying doublet that supports a DO moment. In particular, rare-earth pyrochlores of the form $A_2 B_2 O_7$, where  A and B form a pyrochlore lattice, A is a trivalent rare
earth with a partially filled 4f shell such as $Ce, Nd$, and B is nonmagnetic transition metal such as $Zr, Sn, Sb, Hf$.  These compounds have shown several signatures in experimental probes\cite{DO1,DO2,DO3,DO4,DO5} that indicate a putative QSL state, namely neutron scattering experiments, thermodynamic, magnetometric, and muon relaxation suggests an $U(1)/U(1)_\pi$ QSL ground state in $Ce_2Zr_2 O_7$. Another compound from the same $Ce-$pyrochlore family, $Ce_2Sn_2 O_7$\cite{Ce-spin-liquid}, also has shown a promising signature of QSL ground state.

Our motivation for this present study is the observation of linear temperature dependence (T-linear) of specific heat in $Nd_2 ScNb O_7$\cite{dipole-octapole}. Such  T-linear behavior in a magnetic Mott insulator at low temperatures strongly indicates the presence of deconfined fermionic quasiparticles. Taking a cue from this observation, we perform a  projective symmetry group analysis with fermionic partons and $U(1)$ gauge group for the XYZ model on a pyrochlore lattice where the X and Y spin components transform as dipole moment and Z component transform as a octupole moment. 
Our main results are summarised as follows:
1) restricting to only nearest neighbor terms allowed by the symmetry, we find 12 different $U(1)$ QSL phases (summarised in Table \ref{table_summary}) with non-zero mean-field solutions that are compatible with the space group symmetries of pyrochlore lattice, time-reversal(TR) symmetry and the symmetry transformations of DO magnetic moments. 
2) We find four mean-field states with a unit monopole flux exiting each tetrahedron; such states are dubbed as "monopole-flux" states in literature and known\cite{Burnell} to be a parity and time-reversal breaking chiral phase in $SU(2)$ symmetric spin models on pyrochlore lattice. In the case of DO models, we find "monopole-flux" states to be symmetry-preserving states.  
3) We find four mean-field states to show a T-linear specific heat, the other phases show a power law Temperature dependence of specific heat $C_v \sim T^\alpha$, where $\alpha$ is close to 1. 
4) We compute the dynamic spin structure factors and discuss the possible signature of these fermionic spinons in neutron-scattering experiments on DO magnetic systems.

The rest of this paper is organized as follows. In Sec. II,
we discuss the DO symmetry and the symmetry allowed spin Hamiltonian. In Sec. III we briefly discuss the symmetry group of pyrochlore lattice and introduce the projective symmetry group using parton construction. In Sec. IV, we construct mean-field Hamiltonians for the fermionic spinons and
analyze their symmetry properties. In Sec. V, we study the thermodynamic properties of
the system, such as the specific heat and neutron scattering structure factors. Finally, we summarize and discuss our results in
Sec. VI.
 
 \section{The spin Hamiltonians for dipolar-octupolar pyrochlores}
\label{sec_rto}
\begin{figure}[ht]
\centering
\includegraphics[scale=0.15]{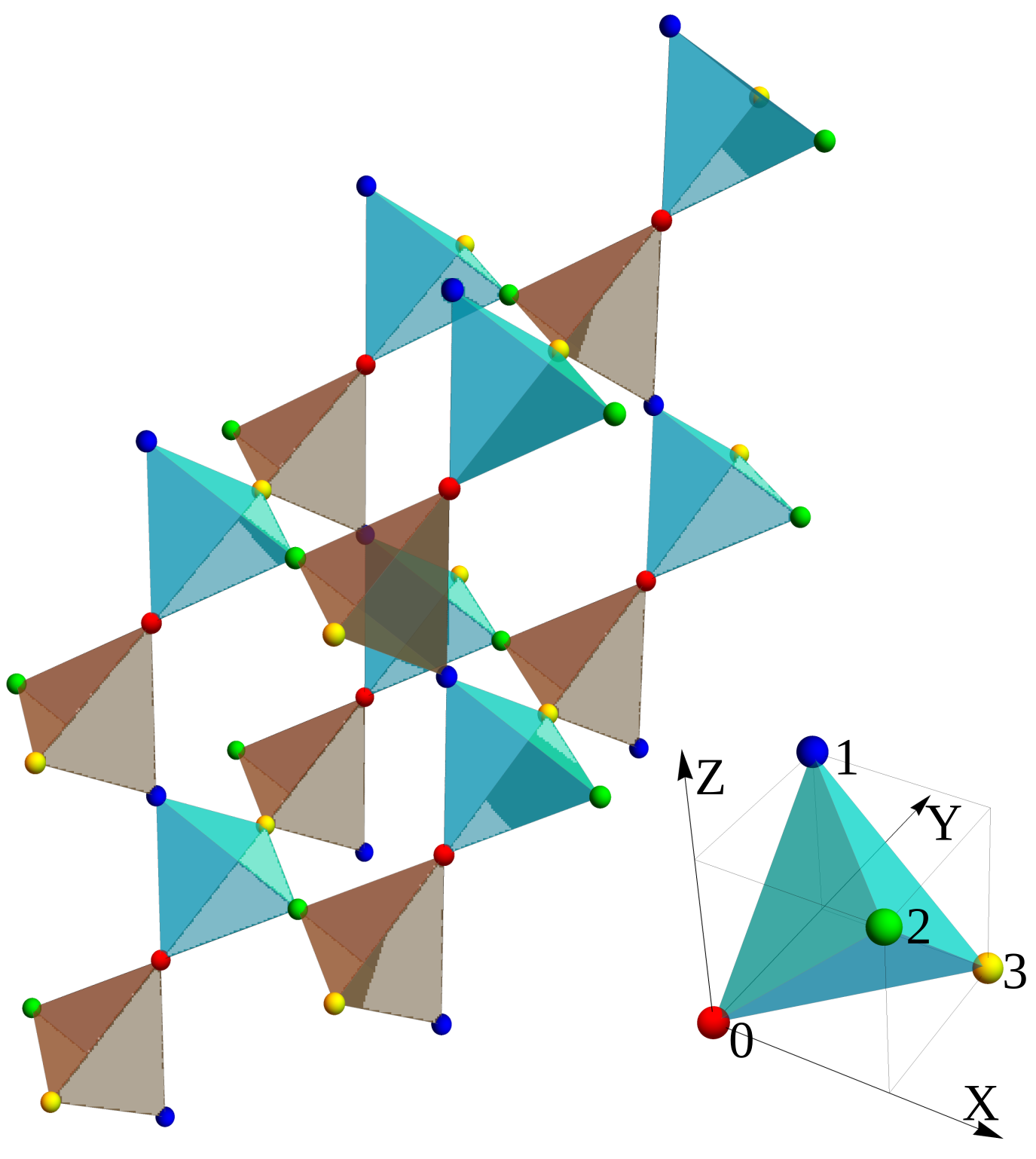}
\caption{(Top) The pyrochlore lattice formed by the rare-earth ions that carry effective spin-$1/2$ moments at each vertex.  (Bottom) The unit cell comprises four sites labeled 0, 1, 2, and 3. The X, Y, and Z represent the global Cartesian directions.}
\label{fig_pyro}
\end{figure}

In this work, we focus on rare-earth pyrochlore magnets with chemical composition A$_2$B$_2$O$_7$ \cite{ROSS, Erbium_pyrochlore, gardner_review}, where both A and B sites form a pyrochlore lattice as shown in Fig. \ref{fig_pyro}. Here B is a transition metal in $d_1$ or $d_3$ electron configuration, A is a trivalent rare-earth with a partially filled $4f$ shell. The magnetism in this system is due to the spin-orbit coupled ${\bf J=L+S}$ magnetic moments on A$^{3+}$. These ${\bf J}$ moments are further split by the local crystal field, giving rise to effective doublets at low energies, which allows for a spin-$1/2$ pseudospin representation. The irreducible representation of the site group of pyrochlore lattice dictates two Kramers and one non-Kramers doublet. The two pseudospin Kramers doublets (as is the case for Yb$_2$Ti$_2$O$_7$ and Er$_2$Sn$_2$O$_7$) and satisfying the usual spin algebra  
$ [s^\mu_i,s^\nu_j]=i\epsilon^{\mu\nu\lambda}s^\lambda_i$
and odd under time-reversal symmetry(TR). The Kramers doublet with the $\Gamma_4$ irreducible representation behaves exactly as a spin-1/2 particle and is dubbed as effective spin 1/2. The other Kramer doublet is constructed from two one-dimensional irreducible representations $\Gamma_5$ and $\Gamma_6$. This doublet is called the dipolar-octupolar (DO) doublet, as $S_i^x$ and $S_i^z$ transform as a dipole and $S_i^y$ transforms as an octupole. This doublet has an interesting property that the pseudospin operators do not mix under the $C_3$ operation. The non-Kramers doublet follows the irreducible representation $E_g$, where $S^z$ is not time-reversal symmetric but $S^\pm$ are time-reversal symmetric.

For the rest of this work, we focus on the pyrochlore with DO symmetry. Due to the trivial action of $C_3$ rotation, the most general nearest neighbor spin interaction model for DO doublet is the simplest spin model with no bond dependence of exchange interaction, the Hamiltonian in the local basis is given by,
\begin{eqnarray}
    H_l &=& \sum_{\langle \mathbf{r}_\mu,\mathbf{r}^{\prime}_\nu \rangle} J_{xx}S_{\mathbf{r}_\mu}^xS_{\mathbf{r}^{\prime}_\nu}^x+J_{yy}S_{\mathbf{r}_\mu}^yS_{\mathbf{r}^{\prime}_\nu}^y+J_{zz}S_{\mathbf{r}_\mu}^zS_{\mathbf{r}^{\prime}_\nu}^z\\ \nonumber
    &+& J_{xz}(S_{\mathbf{r}_\mu}^xS_{\mathbf{r}^{\prime}_\nu}^z+S_{\mathbf{r}_\mu}^zS_{\mathbf{r}^{\prime}_\nu}^x).
 \label{eqn:localH}
\end{eqnarray}
Making a rotation about the local y-axis, i.e. $\hat{y}_\mu$,
\begin{equation*}
    \begin{split}
        & \tau^x = S^x\cos\theta  - S^z\sin\theta \\
        & \tau^y = S^y \\
        & \tau^z = S^x\sin\theta  +S^z \cos\theta ,\\
    \end{split}
\end{equation*}
the Hamiltonian can be simplified as an XXZ model,
\begin{equation}
    H_l = \sum_{a\epsilon\{x,y,z\}} \sum_{\langle \mathbf{r}_\mu,\mathbf{r}^{\prime}_\nu\rangle}\mathcal{J}_a\tau_{\mathbf{r}_\mu}^a\tau_{\mathbf{r}^{\prime}_\nu}^a.
    \label{XXZ_ham}
\end{equation}
Where 
\begin{equation*}
    \begin{split}
    & \tan(2\theta) = \frac{2J_{xz}}{J_{zz}-J_{xx}} \\
    & \mathcal{J}_x = \frac{J_{zz}+J_{xx}}{2}-\frac{\sqrt{(J_{zz}-J_{xx})^2+4J_{xz}^2}}{2} \\
    & \mathcal{J}_y = J_{yy} \\
    & \mathcal{J}_z = \frac{J_{zz}+J_{xx}}{2}+\frac{\sqrt{(J_{zz}-J_{xx})^2+4J_{xz}^2}}{2}. \\ 
    \end{split}
\end{equation*}
In the rest of our PSG analysis, we will work in this $\tau$-basis, and at the end we will restore to the original spin basis for the calculation of dynamical structure factors. The hamiltonian in the global basis and the details of local basis to global basis transformation is given in Appendix \ref{local-global}.

\section{Uniform U(1) QSL with fermionic partons on dipolar-octupolar pyrochlore}
\subsection{Pyrochlore Lattice: space group and time reversal symmetries}
The pyrochlore lattice is made up of an FCC type lattice with four sub-lattices. The sub-lattice sites are denoted by ${\mu}=0,1,2,3$ as in Fig.\ref{fig_pyro}.
The lattice translation vectors of the pyrochlore lattice are given by,
\begin{equation*}
         \textbf{e}_1 = \frac{a}{2}(\hat{y} + \hat{z}) , \textbf{e}_2 = \frac{a}{2}(\hat{z} + \hat{x}), \textbf{e}_3 = \frac{a}{2}(\hat{x} + \hat{y}) .
\end{equation*}
The location of the sublattice sites are given by $\frac{a}{4}(0,0,0),\frac{a}{4}(0,1,1),\frac{a}{4}(1,0,1),\frac{a}{4}(1,1,0)$, where $a$ is the cubic lattice constant. 
We use the notation $(r_1,r_2,r_3)_\mu$ to refer the location of the sub-lattice $\mu$ in the $(r_1,r_2,r_3)$th unit cell, where
\begin{equation*}
     (r_1,r_2,r_3)_\mu = r_1 \textbf{e}_1 + r_2 \textbf{e}_2 +r_3 \textbf{e}_3 + \frac{1}{2} \textbf{e}_{\mu},
\end{equation*}
and $ \textbf{e}_0=(0,0,0)$.

The following five symmetry operations generate the space group of pyrochlore lattice:
    \begin{equation*}
    \begin{split}
       & T_1 : \mathbf{r_\mu} = (r_1,r_2,r_3)_{\mu} \rightarrow (r_1+1,r_2,r_3)_\mu \\
       & T_2 : \mathbf{r_\mu} = (r_1,r_2,r_3)_{\mu} \rightarrow (r_1,r_2+1,r_3)_\mu \\
       & T_3 : \mathbf{r_\mu}= (r_1,r_2,r_3)_{\mu} \rightarrow (r_1,r_2,r_3+1)_\mu \\
       & \Bar{C_6} : \mathbf{r_\mu}= (r_1,r_2,r_3)_{\mu} \rightarrow \\
       &~~~~~~~~~~~~~~~~~ (-r_3-\delta_{\mu,3},-r_1-\delta_{\mu,1},-r_2-\delta_{\mu,2})_{\Bar{C_6}(\mu)} \\
       & S : \mathbf{r_\mu}= (r_1,r_2,r_3)_{\mu} \rightarrow \\
       &~~~~~ (-r_1-\delta_{\mu,1},-r_2-\delta_{\mu,2},r_1+r_2+r_3+1-\delta_{\mu,0})_{S(\mu)}\\
       & \Bar{C_6}(\mu):
       \begin{pmatrix}
       & 0 &\\
       & 1 &\\
       & 2 &\\
       & 3 &\\
       \end{pmatrix}
       \rightarrow
       \begin{pmatrix}
       & 0 &\\
       & 2 &\\
       & 3 &\\
       & 1 &\\
       \end{pmatrix}, ~S(\mu):
       \begin{pmatrix}
       & 0 &\\
       & 1 &\\
       & 2 &\\
       & 3 &\\
       \end{pmatrix}
       \rightarrow
       \begin{pmatrix}
       & 3 &\\
       & 1 &\\
       & 2 &\\
       & 0 &\\
       \end{pmatrix}.
    \end{split}
    \end{equation*}
Where the above symmetry operations are written in the pyrochlore sublattice coordinate system. $T_i$ are the translation operations along the direction $\textbf{e}_i$. Furthermore, $\Bar{C_6}$ is the six-fold rotoinversion around [111] axis, and S is a nonsymmorphic screw operation which is the composition of a  twofold rotation around $\mathbf{e}_3$ and a translation by $\frac{1}{2}\mathbf{e}_3$.

The identity relations between the pyrochlore symmetry group generators are given by:
\begin{equation}
\begin{split}
    T_{i}T_{i+1}T_{i}^{-1}T_{i+1}^{-1}&=1 ,  \forall i
\\
  \Bar{C_6}^6 & = 1\\
  S^2T_3^{-1}& =1\\ \Bar{C_6}T_i\Bar{C_6}^{-1}T_{i+1} & =1\forall i\\
   ST_iS^{-1}T_3^{-1}T_i &=1, i=1,2\\
    ST_3S^{-1}T_3^{-1}&=1\\
    (\Bar{C_6}S)^4 &=1\\
     (\Bar{C_6}^3S)^2&=1\\
     \mathcal{T}^2&=-1\\
\mathcal{T}\mathcal{O}\mathcal{T}^{-1}\mathcal{O}^{-1}&=1,~~~~~\mathcal{O}\in {T_1,T_2,T_3,\Bar{C}_6,S}.
   \end{split}
   \label{null_conditions}
\end{equation}
\subsection{Projective symmetry group}
\label{sec:PSG}
In this section we give an overview of the PSG approach and classify all possible uniform $U(1)$ QSLs on the dipolar-octupolar pyrochlore lattice. We define fermionic spinon/parton annihilation operators as 
$\{f_{\mathbf{r}_\mu\uparrow},f_{\mathbf{r}_\mu\downarrow} \}$ at site $\mathbf{r}_\mu$, and express the $\tau$-spin operators as a bilinear of the fermions as 
\begin{equation}
    \tau_{\mathbf{r}_\mu}^a=\frac{1}{2}f_{\mathbf{r}_\mu\alpha}^\dagger \sigma_{\alpha \beta}^a f_{\mathbf{r}_\mu \beta}, 
    \label{spin-to-spinon}
\end{equation}
where $a \in {1,2,3}$, and $\sigma^a$ refers to the Pauli matrices. Since each spin is represented by two fermions, the Hilbert space is doubled in the spinon representation. For a faithful representation of the Hilbert space, the spinon
operators satisfy the constraint of one spinon per site as
\begin{equation}
\begin{split}
& \sum_\alpha f_{\mathbf{r}_\mu\alpha}^\dagger f_{\mathbf{r}_\mu\alpha}=1,~ {f_{\mathbf{r}_\mu\uparrow}f_{\mathbf{r}_\mu\downarrow}}=0.\\
\end{split}
\label{filling}
\end{equation}
Due to this constraint, the parton description has redundancy that can be manifested as a gauge redundancy. Because of Eq. \ref{filling} one can see that $f_{\mathbf{r}_\mu\uparrow}$ and $f_{\mathbf{r}_\mu\downarrow}^\dagger$ has the same physical effect (manifested as a particle-hole symmetry) i.e. decreasing $\tau_{\mathbf{r}_\mu}^z$ by one unit. Thus, one can place them in a doublet upon which a local $SU(2)$ matrix can act without affecting the physical spin operators. An analogous doublet can be formed by $f_{\mathbf{r}_\mu\downarrow}$, and $-f_{\mathbf{r}_\mu\uparrow}^\dagger$. To manifest this symmetry it is convenient to combine these two doublets and introduce an operator $\Psi_{\mathbf{r}_\mu}=\begin{pmatrix}
                f_{\mathbf{r}_\mu\uparrow} & f_{\mathbf{r}_\mu\downarrow}^{\dagger} \\
                 f_{\mathbf{r}_\mu\downarrow} & -f_{\mathbf{r}_\mu\uparrow}^\dagger\\
             \end{pmatrix}$ .This allow us to re-write Eq. \ref{spin-to-spinon} and \ref{filling} as 
\begin{equation}
\begin{split}
&\tau_{\mathbf{r}_\mu}^a =\frac{1}{4}\text{Tr}[\Psi^\dagger_{\mathbf{r}_\mu} \sigma^a \Psi_{\mathbf{r}_\mu} ]\\
&\text{Tr}(\Psi_{\mathbf{r}_\mu} \sigma^a \Psi^\dagger_{\mathbf{r}_\mu})=0, ~~a ~\in x,y,z
\end{split}
\label{psi_parton}
\end{equation} 
From Eq. \ref{psi_parton} it is easy to see that  spin operators are $SU(2)$ gauge invariant as,
\begin{equation}
    \Psi_{\mathbf{r}_\mu} \rightarrow   \Psi_{\mathbf{r}_\mu} W_{\mathbf{r}_\mu}, \tau_{\mathbf{r}_\mu}^a \rightarrow \tau_{\mathbf{r}_\mu}^a, W_{\mathbf{r}_\mu} \in SU(2).
\end{equation}
As per the definition of QSLs, we are interested in finding the parton Hamiltonians that preserves all the symmetries of the original spin Hamiltonian. One must note that the physical symmetry operators on the spin degrees of freedoms act projectively on the parton operators, which means there could be different parton Hamiltonians that are gauge equivalent. In this case, we want to find all different gauge inequivalent parton Hamiltonians that preserve all the symmetries of the original spin Hamiltonian, the PSG method is a systematic approach to accomplish that task.\\

Under the space group symmetry operation $O$ and TR operation $ \mathcal{T}$, the pseudospin operator transforms as
\begin{equation}
\begin{split}
    & O : \tau_{\mathbf{r}_\mu}^a \rightarrow U_O \tau_{O(\mathbf{r}_\mu)}^a U_{O}^\dagger \\
    & \mathcal{T} : \tau_{\mathbf{r}_\mu}^a \rightarrow \mathcal{K}U_{\mathcal{T}}\tau_{\mathbf{r}_\mu}^a U_{\mathcal{T}}^{\dagger} \mathcal{K}
\end{split}
\end{equation}
where $\mathcal{K}$ is the complex conjugate operator and $U_O$ for the DO doublets are
\begin{equation}
    U_{T_i}=1_{2\times 2}, U_{\Bar{_6}} = 1_{2 \times 2}, U_S = -i \sigma_y, U_{\mathcal{T}}=i \sigma_y,
\end{equation}
a detailed discussion is given in appendix \ref{DO_Symmetry}. 

Due to the $SU(2)$ gauge redundancy any symmetry operation $O$ can be accompanied by a local $SU(2)$ gauge transformation $W_{\mathbf{r}_\mu}$. The projective transformation of the parton operators can be represented as:
\begin{equation}
\begin{split}
    & \widetilde{O}=G_O.O : \Psi_i \rightarrow U_O^\dagger \Psi_{(O(\mathbf{r}_\mu))} W_{O(\mathbf{r}_\mu)}^a \\
    & \widetilde{\mathcal{T}}=G_\mathcal{T}.\mathcal{T} : \Psi_i \rightarrow U_\mathcal{T}^\dagger \Psi_{(\mathcal{T}(\mathbf{r}_\mu))} W_{\mathbf{r}_\mu}^a
    \end{split}
\end{equation}
The $\widetilde{O}=G_O.O$ symbol represents that the projective operation $\widetilde{O}$ combines physical symmetry operation $O$ and the gauge transformation $G_O$. The symmetry group of the projective operations $\{\widetilde{O},\widetilde{\mathcal{T}}\}$ is dubbed as the projective symmetry group (PSG). \\

Our objective is to find all the gauge inequivalent solutions of the projective symmetry group to classify all the symmetry-preserving parton QSL. Those solutions can be obtained from the group identity relations in Eq. \ref{null_conditions}. The group identity relations are implemented modulo its corresponding gauge transformation as 
\begin{equation}
   \widetilde{O}_1.\widetilde{O}_2.\widetilde{O}_3...= (G_{O_1}.O_1).(G_{O_2}.O_2).(G_{O_3}.O_3)...=\mathcal{G},
   \label{null_gauge}
\end{equation}
where $\mathcal{G}$ is a pure gauge transformation modulo of identity transformation. $\mathcal{G}$ is an element of the invariant gauge group (IGG), a subgroup of $\{\widetilde{O},\widetilde{\mathcal{T}},\widetilde{I}\}$. The local IGG transformation is a subgroup of $SU(2)$ i.e. $Z_2$ or $U(1)$, as it is associated with identity operation, it determines the gauge group- thus the classification is often named after it. \\

Using the conjugation relation
\begin{equation}
     O_i. G_{O_i}. O_i^{-1}\rightarrow G_{O_i}[O_i^{-1}]: \Psi_{\textbf{r}_\mu} \rightarrow \Psi_{\textbf{r}_\mu} W_{O_i}[O_i^{-1}(\textbf{r}_\mu)],
\end{equation}
one can express Eq. \ref{null_gauge} purely as a relation between the $SU(2)$ gauge transformation matrices $W$, given by:
\begin{equation}
\begin{split}
  &  W_{O_1}[\textbf{r}_\mu]W_{O_2}[O_1^{-1}(\textbf{r}_\mu)]W_{O_3}[O_2^{-1}O_1^{-1}(\textbf{r}_\mu)]...=\mathcal{G}\\
  &~~~~~~~~~~~~~~~~
  =e^{i\sigma^3 \chi_{O_1 O_2 O_3...}}       
\end{split}
\end{equation}
For $U(1)$ IGG, $\chi_{O_1 O_2 O_3...} \in [0,2\pi)$, which constraints the general form of $W_O(\textbf{r})=(i \sigma^1)^{w_{O_i}} e^{i\sigma^3 \phi_{O_i}(\textbf{r})}$. Due to the properties of Pauli matrices, only the parity of $w_{O_i}$ matters i.e. $w_{O_i}$ is either $0$ or $1$. The parity nature of $w_{O_i}$, will further quantize allowed values of some of the $\chi$ parameters, and through Eq. \ref{null_gauge}  the $\phi$ parameters will be determined by $\chi$. Therefore even though $U(1)$ IGG allows the $\chi_{O_1 O_2 O_3...}$ to be a continuous variable, the number of possible PSG solutions with $U(1)$ IGG will be finite.

The relations Eq. \ref{null_gauge} further constrainsts $w_{T_i}=0$ and we are left with $w_{\bar{C}_6}$, $w_S$, $w_{\mathcal{T}}$ and the $\phi$ parameters. For different choices of $w$'s one can use the relations Eq. \ref{null_gauge} to find a relation between the $\phi$ parameters and $\chi$ parameters. Notably the Eq. \ref{null_gauge} will give a difference equation in phase parameter $\phi$ that will require a further gauge choice to fix the 'origin' in the phase $\phi$, we choose this as $\phi_{T_1}(\mathbf{r}_\mu)=0$, $\phi_{T_2}(\mathbf{r}_\mu)=-\chi_1 r_1$, $\phi_{T_3}(\mathbf{r}_\mu)=\chi_3 r_1 -\chi_2 r_2$. The detailed study of the PSG equations with U(1) IGG are given in \cite{Liu_fermion}. 

Since our main objective is to find all the different $W$s that are unrelated by a gauge transformation, we need to also consider the gauge equivalent group relations,

\begin{equation}
  (G.G_{O_1}.O_1.G^{-1}).(G.G_{O_2}.O_2.G^{-1}).(G.G_{O_3}.O_3.G^{-1})...=\mathcal{G},
   \label{null_gauge_}
\end{equation}
that translates to a relation between the $SU(2)$ gauge transformations $W$ as 
\begin{equation}
      W_{O_i}(\textbf{r}_\mu) \rightarrow W(\textbf{r}_\mu) W_{O_i}(\textbf{r}_\mu) W^{-1}[{O_i}(\textbf{r}_\mu)].
\end{equation}
We need to ensure that our PSG solutions are distinct modulo this gauge transformation. We do a gauge fixing by taking $W(\textbf{r}_0)=I$ and $W(\textbf{r}_i)=e^{i\sigma_3 \psi_i r_i} , i = 1,2,3 $. Under this gauge fixing one must fix $\chi_{\bar{C}_6 T_1}=\chi_{\bar{C}_6 T_2}=\chi_{S T_2}=0$ to keep the PSG solutions invariant. Imposing the TR symmetry relations further sets some more $\chi$ and $w$ parameters to a fixed value. Finally, the full PSG solutions can be expressed in terms of only 4 free-parameters that are $Z_2$ variables, namely $\chi_1, \chi_{\bar{C}_6 S},w_{\bar{C}_6 }, w_S$. The final PSG solutions look like:
\begin{equation}
    \begin{split}
        & W_{T_i}(\mathbf{r}_\mu)= e^{i\sigma^{3}\phi_{T_i}(\mathbf{r}_\mu)},     i=1,2,3 \\
        & W_{\bar{C}_6}(\mathbf{r}_\mu)= W_{\bar{C}_6,\mu}e^{i\sigma^3\phi_{\bar{C}_6}(\mathbf{r}_\mu)} \\
        & W_S(\mathbf{r}_\mu) = W_{S,\mu}e^{i\sigma^3\phi_{S}(\mathbf{r}_\mu)} \\
        & W_{\mathcal{T}}(\mathbf{r}_\mu)=i\sigma^1 \\
    \end{split}
\end{equation}
Where the values of $W_{\bar{C}_6,\mu}, W_{S,\mu}$ for different values of $\chi_1,w_{\Bar{C}_6},w_S,\chi_{ST_1}$ and $\chi_{\Bar{C}_6S}$ are given in table \ref{full psg classification}. The phase $\phi(\mathbf{r}_\mu)$ for each of the symmetry are given by, 
\begin{equation}
    \begin{split}
        & \phi_{T_1}(\mathbf{r}_\mu) = 0, \\
        & \phi_{T_2}(\mathbf{r}_\mu) = -\chi_1r_1, \\
        & \phi_{T_3}(\mathbf{r}_\mu) = \chi_1(r_1-r_2), \\
        & \phi_{\bar{C}_6}(\mathbf{r}_\mu) = -\chi_1r_1(r_2-r_3)+\delta_{\mu,2}\chi_1r_3\\
        &~~~~~~~~~~~~~-r_1[2\chi_{ST_1}+2\chi_1+(\delta_{\mu,2}-\delta_{\mu,3})\chi_1],\\ 
        & \phi_{S}(\mathbf{r}_\mu) = \chi_1[\frac{r_1(r_1+1)}{2}-\frac{r_2(r_2+1)}{2}-r_1r_2]\\
        &~~~~~~~~~~~~~+r_1[\chi_1(\delta_{\mu,1}-\delta_{\mu,2})+(2\chi_1-\chi_{ST_1})]\\
        &+r_2[(2\delta_{\mu,1}-\delta_{\mu,2})\chi_1+3\chi_{ST_1}]+\chi_1r_3[(\delta_{\mu,1}-\delta_{\mu,2})+2]
    \end{split}
\end{equation}

\begingroup
\setlength{\tabcolsep}{8 pt}
\renewcommand{\arraystretch}{1.3} 
\begin{table*}
\centering
    \begin{tabular}{c c c c c c c c}
    \hline
    $\chi_1$ & $\chi_{\bar{C}_6S}$ & $w_{\bar{C}_6}$ & $w_S$ & $(W_{\bar{C}_6,0},W_{\bar{C}_6,1},W_{\bar{C}_6,2},W_{\bar{C}_6,3})$ & $(W_{S,0},W_{S,1},W_{S,2},W_{S,3})$ & Classes \\
    \hline
    0,$\pi$  & 0,$\pi$ & 0 & 0 & $(1,1,1,e^{i\sigma^3(\chi_1-\chi_{ST_1})})$ & $(1,e^{i\sigma^3(\chi_{ST_1}+\chi_{\bar{C}_6S})},1,e^{i\sigma^3\chi_1})$ & 4 \\
    
    0,$\pi$  & 0,$\pi$ & 0 & 1 & $(1,1,1,e^{i\sigma^3(\chi_1-\chi_{ST_1})})$ & $(i\sigma^1,i\sigma^1e^{i\sigma^3(\chi_{ST_1}+\chi_{\bar{C}_6S})},i\sigma^1,i\sigma^1e^{i\sigma^3\chi_1})$ & 4 \\
    
    0,$\pi$  & 0,$\pi$ & 1 & 0 & $(i\sigma^1,i\sigma^1,i\sigma^1,i\sigma^1e^{i\sigma^3(\chi_1-\chi_{ST_1})})$ & $(1,e^{i\sigma^3(\chi_{ST_1}+\chi_{\bar{C}_6S})},1,e^{i\sigma^3\chi_1})$ & 4 \\
    
    0,$\pi$ & 0,$\pi$ & 1 & 1 & $(i\sigma^1,i\sigma^1,i\sigma^1,i\sigma^1e^{i\sigma^3(\chi_1-\chi_{ST_1})})$ & $(i\sigma^1,i\sigma^1e^{i\sigma^3(\chi_{ST_1}+\chi_{\bar{C}_6S})},i\sigma^1,i\sigma^1e^{i\sigma^3\chi_1})$ & 4 \\
    \hline
    \end{tabular}
    \caption{This table enlist the form of $W_{\bar{C}_6,\mu}$, $W_{S,\mu}$ matrices for 16 TR symmetric U(1) PSG classes. Each of the PSG class is parametrized with $\chi_1$, $\chi_{ST_1}$, $\chi_{C_6}$, $w_{C_6}$ and $w_S$ where $\chi_{ST_1}=0$ for all the classes.}
    \label{full psg classification}
    \end{table*}
\endgroup
Different PSG classes described by different values of $\chi_1,w_{\Bar{C}_6},w_S,\chi_{ST_1}$ and $\chi_{\Bar{C}_6S}$ which is given in the table \ref{naming-psg}.
\begingroup
\setlength{\tabcolsep}{6 pt}
\renewcommand{\arraystretch}{1.3} 
\begin{center}
\begin{table}
    \begin{tabular}{c c c c c c c}
       Case No.  & $\chi_1$ & $\chi_{ST_1}$ & $\chi_{C_6}$ & $w_{C_6}$  & $w_S$  \\
       \hline
        1        & 0       &  0            & 0            &  0         & 0      \\
        2        & $\pi$       &  0            & 0            &  0         & 0      \\
        3        & 0       &  0            & $\pi$            &  0         & 0      \\
        4        & $\pi$       &  0            & $\pi$            &  0         & 0   \\
        5        & 0       &  0            & 0            &  0         & 1      \\
        6        & $\pi$       &  0            & 0            &  0         & 1      \\
        7        & 0       &  0            & $\pi$            &  0         & 1      \\
        8        & $\pi$       &  0            & $\pi$            &  0         & 1   \\
        9        & 0       &  0            & 0            &  1         & 0      \\
        10        & $\pi$       &  0            & 0            &  1         & 0      \\
        11        & 0       &  0            & $\pi$            &  1         & 0      \\
        12        & $\pi$       &  0            & $\pi$            &  1         & 0   \\
        13       & 0       &  0            & 0            &  1         & 1      \\
        14        & $\pi$       &  0            & 0            &  1         & 1      \\
        15        & 0       &  0            & $\pi$            &  1         & 1      \\
        16        & $\pi$       &  0            & $\pi$            &  1         & 1   \\
        \hline
    \end{tabular}
    \caption{We identify different PSG classes with different case numbers which are characterized by $\chi$'s and $w$'s listed in the table. We are using the above convention to refer to different PSG classes.}
    \label{naming-psg}
\end{table}
\end{center}
\endgroup
\section{Parton mean field theory}
In the previous section, we showed the PSG classification of all the fractionalized parton representations of pseudospins that preserves all the space group symmetries on the pyrochlore lattice and the TR symmetry. Notably, the transformation properties of pseudospins itself doesn't matter here due to the conjugation relation (Eq. 14), which comes from the null conditions and are same for both dipolar-octupolar and spin-1/2, so this PSG classification remains the same as that of Kramer's spin . However, using the parton representation in Eq. \ref{spin-to-spinon} we have transformed the bilocal pseudospin Hamiltonian in Eq. \ref{XXZ_ham} to a quartic spinon/parton Hamiltonian, that reads as:
\begin{equation}
    \begin{split}
       & \mathcal{J}_x \tau_i^x 
    \tau_j^x =-\frac{\mathcal{J}_x}{4}\sum_{\alpha}f_{i\alpha}^\dagger f_{j\alpha}^\dagger f_{i
\bar{\alpha}}f_{j\bar{\alpha}}+ f_{i\alpha}^\dagger f_{j\bar{\alpha}}^\dagger f_{i\alpha
}f_{j\bar{\alpha}}\\
       & \mathcal{J}_y \tau_i^y 
    \tau_j^y  =\frac{\mathcal{J}_y}{4}\sum_{\alpha}(f_{i\alpha}^\dagger f_{j\alpha}^\dagger f_{i
\bar{\alpha}}f_{j\bar{\alpha}} -f_{i\alpha}^\dagger f_{j\bar{\alpha}}^\dagger f_{i\bar{\alpha}}f_{j\alpha})\\
       & \mathcal{J}_z \tau_i^z
    \tau_j^z=-\frac{\mathcal{J}_z}{4}\sum_{\alpha}(f_{i\alpha}^\dagger f_{j\alpha}^\dagger f_{i
\alpha}f_{j\alpha} -f_{i\alpha}^\dagger f_{j\bar{\alpha}}^\dagger f_{i\alpha}f_{j\bar{\alpha}}),\\
    \end{split}
\end{equation}
where $\{\alpha,\beta\}=\{\uparrow,\downarrow\}.$ In order to perform a mean-field theory with symmetry preserving "order parameters" one needs to factorize the quartic fermionic operators in terms of all possible symmetric quadratic operators, which are given by:
\begin{equation}\label{mf_channels}
  \begin{split} 
     & \hat{\chi}_{ij}=f_{i\alpha}^\dagger \delta_{\alpha\beta}f_{j\beta}, ~\hat{\Delta}_{ij}= f_{i\alpha}[i\sigma^y]_{\alpha\beta}f_{j\beta}  \\
     & \hat{E}_{ij}^{a}=f_{i\alpha}^{\dagger}[\sigma^a]_{\alpha\beta}f_{j\beta},~\hat{D}_{ij}^a=f_{i\alpha}[i\sigma^y\sigma^a]_{\alpha\beta}f_{j\beta} \\
     & \hat{n}_i=f_{i\alpha}^\dagger \delta_{\alpha\beta}f_{i\beta}.\\
  \end{split}    
\end{equation}
The spin Hamiltonian in Eq. \ref{XXZ_ham}, in the isotropic limit i.e. $\mathcal{J}_x=\mathcal{J}_y=\mathcal{J}_z$, can be factorized in terms of the symmetric quadratic operators (Eq. \ref{mf_channels}) as,
\begin{equation}
    \begin{split}
    H=&J\sum_{\langle i,j \rangle}-\frac{3}{4}(\hat{\chi}_{ij}^\dagger \hat{\chi}_{ij}+\hat{\Delta}_{ij}^\dagger \hat{\Delta}_{ij} +{\hat{E}_{ij}}^{x\dagger} \hat{E}_{ij}^x+{\hat{E}_{ij}}^{y\dagger} \hat{E}_{ij}^y\\
    & +{\hat{E}_{ij}}^{z\dagger} \hat{E}_{ij}^z)-\frac{1}{4}({\hat{D}_{ij}}^{x\dagger} \hat{D}_{ij}^x+{\hat{D}_{ij}}^{y\dagger} \hat{D}_{ij}^y+{\hat{D}_{ij}}^{z\dagger} \hat{D}_{ij}^z)\\
    &-\frac{3}{4}\hat{n}_i\hat{n}_j+3\hat{n}_j.\\
 \end{split}
\end{equation}
A mean-field Hamiltonian is achieved by replacing the mean-field channels with its ground state expectation value, i.e., $\hat{A}_{ij}=A_{ij}+\delta A_{ij}$, where $\hat{A}$ are the quadratic fermionic operators and $\delta A$ are the fluctuations about the expectation value $A_{ij}$. The corresponding mean-field Hamiltonian with  first order in fluctuation of mean-field is given by,
\begin{equation}\label{generic_mf}
    \begin{split}
    H_{MF}=&J\sum_{\langle i,j \rangle}-\frac{3}{4}({\chi}_{ij}^* \hat{\chi}_{ij}+{\Delta}_{ij}^* \hat{\Delta}_{ij} +{{E}_{ij}^x}^* \hat{E}_{ij}^x+{{E}_{ij}^y}^* \hat{E}_{ij}^y\\
    &+{{E}_{ij}^z}^* \hat{E}_{ij}^z)-\frac{1}{4}({{D}_{ij}^x}^* \hat{D}_{ij}^x+{{D}_{ij}^y}^* \hat{D}_{ij}^y+{{D}_{ij}^z}^* \hat{D}_{ij}^z)\\
    &+\text{h.c.}+ \frac{3}{4}(\abs{\chi_{ij}}^2+\abs{\Delta_{ij}}^2+\abs{E^x_{ij}}^2+\abs{E^y_{ij}}^2\\
    &+\abs{E^z_{ij}}^2)
    +\frac{1}{4}(\abs{D^x_{ij}}^2+\abs{D^y_{ij}}^2+\abs{D^z_{ij}}^2)\}]\\
    \end{split}
\end{equation}
A detailed derivation of Eq. \ref{generic_mf} is given in appendix \ref{generic mf decomposition}.

\subsection{Construction of the mean-field Ansätze}
Eq. \ref{generic_mf} is one of the generic mean-field decompositions of the spin Hamiltonian. However, projective symmetry determines the allowed mean-field process for each PSG class. One can construct the mean-field Hamiltonian for each PSG class to understand that. The linear terms in mean-field operators in Eq. \ref{generic_mf} can be written in $SU(2)$ basis,

\begin{eqnarray}
\tilde{H}&=&\sum_{\alpha=0,x,y,z} \tilde{H}^\alpha \\ \nonumber
&=& -\frac{3}{4}J\sum_{\alpha=0,x,y,z}\sum_{\mathbf{r}_{\mu},\mathbf{r}_{\nu}^{\prime}}\text{Tr}[\sigma^{\alpha}\Psi_{\mathbf{r}_{\mu}}u^{\alpha}_{\mathbf{r}_{\mu},\mathbf{r}_{\nu}^{\prime}}\Psi_{\mathbf{r}_{\nu}^{\prime}}^{\dagger}],
\end{eqnarray}
where the mean-field $u_{ij}^{\alpha}$ matrices are written as
\begin{center}

    \begin{tabular}{c c}
        $
             u_{\mathbf{r}_{\mu},\mathbf{r}_{\nu}^{\prime}}^{0}=\begin{pmatrix}
                 -\chi_{\mathbf{r}_{\mu},\mathbf{r}_{\nu}^{\prime}} & \Delta_{\mathbf{r}_{\mu},\mathbf{r}_{\nu}^{\prime}}^{*} \\
                 \Delta_{\mathbf{r}_{\mu},\mathbf{r}_{\nu}^{\prime}} & \chi_{\mathbf{r}_{\mu},\mathbf{r}_{\nu}^{\prime}}^{*} \\
             \end{pmatrix}
         $,
         \\ 
         \\
          $
             u_{\mathbf{r}_{\mu},\mathbf{r}_{\nu}^{\prime}}^{a}=\begin{pmatrix}
                 -{E^{a}}_{\mathbf{r}_{\mu},{\mathbf{r}}^{\prime}_{\nu}} & -3{D^{a}}^*_{\mathbf{r}_{\mu},\mathbf{r}_{\nu}^{\prime}} \\
                 3{D^{a}}_{\mathbf{r}_{\mu},\mathbf{r}_{\nu}^{\prime}} & -{E^{a}}^*_{\mathbf{r}_{\mu},\mathbf{r}_{\nu}^{\prime}} \\
             \end{pmatrix},
         $\\
    \end{tabular}
\end{center}
where $a={x,y,z}$, and $\chi_{\mathbf{r}_{\mu},\mathbf{r}_{\nu}^{\prime}}, \Delta_{\mathbf{r}_{\mu},\mathbf{r}_{\nu}^{\prime}}, E_{\mathbf{r}_{\mu},\mathbf{r}_{\nu}^{\prime}}^a, D_{\mathbf{r}_{\mu},\mathbf{r}_{\nu}^{\prime}}^a$ are the complex mean-field parameters. Under U(1) IGG
\begin{equation}
\begin{split}
H & = \sum_{\alpha = 0,x,y,z}\sum_{\mathbf{r}_{\mu},\mathbf{r}_{\nu}^{\prime}} \text{Tr}[\sigma^\alpha\Psi_{\mathbf{r}_{\mu}} u_{\mathbf{r}_{\mu},\mathbf{r}_{\nu}^{\prime}}^{\alpha}\Psi_{\mathbf{r}_{\nu}^{\prime}}^{\dagger}]\\
& \rightarrow \sum_{\alpha = 0,x,y,z}\sum_{\mathbf{r}_{\mu},\mathbf{r}_{\nu}^{\prime}} \text{Tr}[\sigma^\alpha\Psi_{\mathbf{r}_{\mu}}W u_{\mathbf{r}_{\mu},\mathbf{r}_{\nu}^{\prime}}^{\alpha}W^\dagger \Psi_{\mathbf{r}_{\nu}^{\prime}}^{\dagger}],
\end{split}
\end{equation}
where 
\begin{equation*}
    W = e^{i\sigma^z\phi}, \phi~\epsilon~(0,2\pi).
\end{equation*}
Since it is a gauge operation, expressions on both side of the arrow must be equal. This makes the $\Delta_{\mathbf{r}_{\mu},\mathbf{r}_{\nu}^{\prime}},D_{\mathbf{r}_{\mu},\mathbf{r}_{\nu}^{\prime}}^{a}$ zero. Now under TR symmetry,
\begin{equation}
\begin{split}
    &
    u_{\mathbf{r}_{\mu},\mathbf{r}_{\nu}^{\prime}}^{\alpha}=-W_{\mathcal{T}}(\mathbf{r}_{\mu}) u_{\mathbf{r}_{\mu},\mathbf{r}_{\nu}^{\prime}}^{\alpha} W_{\mathcal{T}}^\dagger(\mathbf{r}_{\nu}^{\prime}),\\
    \end{split}
\end{equation}
where
\begin{equation*}
    U_{\mathcal{T}} = i\sigma^y~,~ W_{\mathcal{T}}(i)=i\sigma^x .
\end{equation*}
This requires $\chi_{\mathbf{r}_{\mu},\mathbf{r}_{\nu}^{\prime}}$, to be pure real and $ E_{\mathbf{r}_{\mu},\mathbf{r}_{\nu}^{\prime}}^a$ to be pure imaginary. Now we left with,
\begin{center}
    \begin{tabular}{c c}
        $
             u_{\mathbf{r}_{\mu},\mathbf{r}_{\nu}^{\prime}}^{0}=\begin{pmatrix}
                 -\chi_{\mathbf{r}_{\mu},\mathbf{r}_{\nu}^{\prime}} & 0 \\
                 0 & \chi_{\mathbf{r}_{\mu},\mathbf{r}_{\nu}^{\prime}} \\
             \end{pmatrix}
         $,
         &  
          $
             u_{\mathbf{r}_{\mu},\mathbf{r}_{\nu}^{\prime}}^{a}=\begin{pmatrix}
                 -E_{\mathbf{r}_{\mu},\mathbf{r}_{\nu}^{\prime}}^{a} & 0 \\
                 0 & {E_{\mathbf{r}_{\mu},\mathbf{r}_{\nu}^{\prime}}^{a}}\\
             \end{pmatrix}
         $\\
    \end{tabular}
\end{center}
Therefore, with U(1) IGG and TR PSG, equation(21) reduces to
\begin{equation}\label{mf_chi_e}
    \begin{split}
    H_{MF}=&J\sum_{\langle i,j \rangle}-\frac{3}{4}({\chi}_{ij} \hat{\chi}_{ij}+{{E}_{ij}^x}^* \hat{E}_{ij}^x+{{E}_{ij}^y}^* \hat{E}_{ij}^y
    +{{E}_{ij}^z}^* \hat{E}_{ij}^z)\\
    &+\text{h.c.}+ \frac{3}{4}(\abs{\chi_{ij}}^2+\abs{E^x_{ij}}^2+\abs{E^y_{ij}}^2+\abs{E^z_{ij}}^2)]\\
    \end{split}
\end{equation}
Although the allowed mean-field channels are confined to $\chi_{ij}$ and $E^a_{ij}$, still it is difficult to solve as there will be as many $\chi$'s and $E^a$'s as many bonds are there. The problem can be further simplified by using the symmetries of the pyrochlore space group to establish the relation between the mean field on different bonds. The key idea is, that upon a symmetry transformation, one bond gets mapped to another bond, which in turn puts a constraint between the mean-field on those two bonds. We have given a detailed procedure for finding the constraints among the nearest neighbor mean field in appendix \ref{mf symmetric relations}. 

We found that for PSG class 5, 6, 9, and 10 all the nearest neighbor mean field vanishes. For case 1, case 2, case 13, and case 14 the mean-field Hamiltonian is
\begin{equation*}
 H_{MF}=-\frac{3}{4}J\left(\sum_{\langle \mathbf{r}_{\mu},\mathbf{r}_{\nu}^{\prime}\rangle}{\chi}_{\mathbf{r}_{\mu},\mathbf{r}_{\nu}^{\prime}}\hat{\chi}_{\mathbf{r}_{\mu},\mathbf{r}_{\nu}^{\prime}}+h.c.\right)+\frac{9}{4}JN\chi^2 
\end{equation*}
Similarly, for PSG cases 3, 4, 15, and 16 
\begin{equation*}
\begin{split}
    &H_{MF}=-\frac{3}{4}J\left(\sum_{\langle i,j \rangle}{{E}_{ij}^x}^* \hat{E}_{ij}^x
    +{{E}_{ij}^z}^* \hat{E}_{ij}^z+\text{h.c.} \right) \\
    & ~~~~~~~~~+ \frac{3}{4}JN(\abs{E^x}^2+\abs{E^z}^2)]\\
\end{split}
\end{equation*}
For PSG cases 7, 8, 11, and 12 the mean-field Hamiltonian will be,
\begin{equation*}
    H_{MF}=-\frac{3}{4}J\left(\sum_{\langle i,j \rangle}{{E}_{ij}^y}^* \hat{E}_{ij}^y
    +\text{h.c.}\right)+ \frac{9}{4}JN\abs{E^y}^2
\end{equation*}

Although a group of PSG class follows similar Hamiltonian, but the constraints on the mean-fields are different. These constraints are given in Appendix \ref{mf symmetric relations}.
\subsection{Minimization}
The mean-field Hamiltonian is a function of mean-field parameters. $H_{MF}=H_{MF}({A})$, where are $A$'s are the mean-field parameters. This mean-field Hamiltonian is variationally solved while satisfying the half-filling constraint given in equation(\ref{filling}). 
We minimize
\begin{equation}
    \begin{split}
        \frac{E_{MF}}{N} = & \sum_{A}c^A\abs{A}^2 \\
        & +\frac{1}{4N}\sum_{k_x,k_y,k_z \epsilon \text{BZ}}\sum_{i = 1}^{N_{band}}A\Lambda_i\left(\mathbf{k},J\right)\Theta\left(\mu,\Lambda_i\left(\mathbf{k},J\right) \right)
    \end{split}
\end{equation}
Where $\Lambda(\mathbf{k},J)$ is the energy dispersion of the partons and $\Theta(\mu,\Lambda_i\left(\mathbf{k},J\right))$ is the fermi function at zero temperature. The chemical potential $\mu$ is set by the single occupancy constraint,
\begin{equation}
\frac{1}{N_{band}}\frac{2}{N_{UC}}\sum_{k_x k_y k_z \in BZ} ~\sum_{i=1}^{N_{band}}~\Theta(\mu,\Lambda_i(\textbf{k},J))=1.
\end{equation}.

Table \ref{table_summary} summarises the nearest neighbor mean-field result. Out of the 16 time reversal PSG classes 4 PSG classes do not allow any nearest neighbor mean-field solution. For the remaining 12 PSG classes, we study the specific heat and structure factors in the next sections.
\begin{center}
\begin{table*}[hbt!]

    \begin{tabular}{c c c c c c c c c}
    \hline
    Case No. &  Constraints & Non-Zero &Energy Dispersion & $E_{min}$ & $C_V~\alpha~T^x$ & Flux State of& Flux State of\\
            &               &        Parameters  &Near Fermi level  &                    &          &  Down-pointing & Up-pointing \\
     & & & & & &Tetrahedra &Tetrahedra\\
     
    1 & $\Delta=E^a=D^a=0$ & $\chi$ &Gapless at all k& $-0.25J$ & Upturn & 0 & 0\\

     &  &  &near Fermi energy&  &  &  & \\
    
    2 & $\Delta=E^a=D^a=0$ & $\chi$& &  $-0.251J$ & Upturn & $\pi$ & $\pi$ \\

    &  &  &  &  &  &  & \\
    
    3 & $\Delta=E^y=D^a=\chi=0$ & $E^x$,$E^z$&Gapless along W$\rightarrow$L & $-0.2989J$ & $x\sim 1.04$ & $-\frac{\pi}{2}$ & $\frac{\pi}{2}$ \\

    &  &  &Linear Gapless near K&  &  &  & \\
    
    4 &$\Delta=E^y=D^a=\chi=0$ & $E^x$,$E^z$ &-- & $-0.317J$ & $x\sim$ 0.90& $\frac{\pi}{2}$ & $-\frac{\pi}{2}$\\

    5 & All the MF channels are zero & - & - & - \\
    
    6 & All the MF channels are zero & - & - & - \\
    
    7 & $\Delta=E^x=E^z=D^a=\chi=0$ & $E^y$&Gapless along W$\rightarrow$L  & $-0.2988$ & $x\sim0.996$ & ($0,\pi,\pi,0$) &($\pi,0,0,\pi$) \\
     &  &  &Linear Gapless near K&  &  &  & \\
    8 & $\Delta=E^x=E^z=D^a=\chi=0$ & $E^y$&-- & $-0.2683J $& $x\sim0.84$ & ($0,\pi,\pi,0$) &($\pi,0,0,\pi$) \\ 
    
    9 & All the MF channels are zero & - & -\\
    
    10 & All the MF channels are zero & - & - \\
    
    11 & $\Delta=E^x=E^z=D^a=\chi=0$ & $E^y$ & Gapless along L$\rightarrow\Gamma$ & $-0.3148J$ & $x\sim 1.27$ & ($0,\pi,\pi,0$) &($0,\pi,\pi,0$)\\
     &  &  &Linear Gapless at $\Gamma$&  &  &  & \\
    
    12 & $\Delta=E^x=E^z=D^a=\chi=0$ & $E^y$ &-- & $-0.266J$ & $x\sim0.96$ &  ($\pi,0,0,\pi$) &($\pi,0,0,\pi$) \\

    13 & $\Delta=E^a=D^a=0$ & $\chi$ &Gapless at all k & $-0.1820J$ & $x\sim 1.15 $ & $0$ & $\pi$\\
     &  &  &Linear Gapless at $\Gamma$&  &  &  & \\
    
    14 & $\Delta=E^a=D^a=0$ & $\chi$ &-- & $-0.1196J$ & $x\sim1.15$ &   $\pi$ & $0$\\

    15 & $\Delta=E^y=D^a=\chi=0$ & $E^x$,$E^z$ & Gapless along L$\rightarrow\Gamma$ & $-0.3148J$ & $x\sim 1.27$ &  $\frac{\pi}{2}$ & $\frac{\pi}{2}$\\
     &  &  &Linear Gapless at $\Gamma$&  &  &  & \\
    
    16 & $\Delta=E^y=D^a=\chi=0$ & $E^x$,$E^z$ &-- & $-0.2971J$ &  $x\sim0.95$ &$\frac{\pi}{2}$ & $\frac{\pi}{2}$\\
    \hline
    
    \end{tabular}
    \caption{This table summarizes the ground state energies, corresponding specific heat behavior with temperature, and the flux state of the pyrochlore unit cell for all the PSG classes mentioned above. It is important to note that the odd-numbered cases have the unit cell of a regular pyrochlore with 4 sub-lattice and 12 bonds (One down-pointing and one up-pointing tetrahedra), however, the even-numbered cases have a unit cell which is four times larger than the regular pyrochlore unit cell with 16 sub-lattices and 48 bonds (4 down-pointing and 4 up-pointing tetrahedra). The last two columns detail the flux through each of the triangular faces of down-pointing and up-pointing tetrahedra. Note that cases 7, 8, 11, and 12 do not bind the same flux through all of their triangular faces of tetrahedra. There are 4 entries in both the columns denoting the flux state of down-pointing and up-pointing tetrahedra. The 4 entries represent flux through 4 triangular surfaces spanned by the sub-lattices 0,1,2,3. The triangular surfaces in down-pointing tetrahedra are 0-2-1-0, 0-1-3-0, 1-2-3-1, 0-3-2-0 and in up-pointing tetrahedra are 0-3-1-0, 0-1-2-0, 0-2-3-0, 1-3-2-1. The flux bindings are written in the same order in which the surfaces are written.}

\label{table_summary}
\end{table*}
\end{center}
\section{Experimentally measurable signatures of the parton mean field ansatzs}
\subsection{Spinon Band Structure}
Here we discuss the energy dispersion of the spinons for the 12 PSG classes. The band structure for cases 1, 3, 11, and 13 are given in Fig. \ref{dispersion_spheat_fig} for which the NN mean fields are non-zero. For PSG case 1, the energy dispersion is the same as the tight-binding model on a pyrochlore lattice which is a zero flux state. It has a flat band with two-fold degeneracy at the Fermi level. The PSG case 3 binds a monopole flux in both the tetrahedra through the $\hat{E}^x$ and $\hat{E}^z$ channels. It has a linear dispersion near the K point. The band structure of cases 3 and 7 are identical although the allowed mean fields for case 3 are $\hat{E}^x$ and $\hat{E}^z$, and for case 7 are $\hat{E}^y$. Similarly, PSG cases 15 and 11 have identical energy dispersion; however, the allowed mean-field channels in case 15 are $\hat{E}^x$ and $\hat{E}^z$, but for case 11, it is $\hat{E}^y$. The PSG case 13 shows a flat band with two-fold degeneracy near the Fermi energy with linear dispersion at $\Gamma$ point. 
\subsection{Specific Heat}
In this section, we study the temperature dependence of specific heat which can be measured in the experiment. Schei et al. \cite{C_Broholm} reported linear T behavior in $\text{Nd}_2\text{ScNbO}_7$ at low temperatures. As this compound is a Mott insulator, one plausible explanation for the linear T behavior is the fermionic spin liquid excitations. Moreover, the low-lying states of this compound have dipolar-octupolar symmetry. The temperature dependence of specific heat could be a signature for the spin liquid states. 

Taking motivation from this experiment, we calculated the specific heat for all the PSG classes. We evaluate,
\begin{equation}
\begin{split}
    C_V&=\frac{\partial E}{\partial T}\\
    &=\sum_{\mathbf{k}\in \text{BZ},\alpha}g\epsilon_\alpha(\mathbf{k})\frac{\epsilon_\alpha(\mathbf{k})-\mu}{k_BT^2} f^2(\epsilon_\alpha(\mathbf{k}),\mu,T)e^{\frac{\epsilon_\alpha(\mathbf{k})-\mu}{k_BT}}
\end{split}
\end{equation}
where E is the total energy of the system at half-filling and g is degeneracy of spinon bands $\epsilon_\alpha(\mathbf{k})$ indexed with $\alpha$ and $f(\epsilon_\alpha(\mathbf{k}),\mu,T)$ is fermi distribution and $\mu$ is the fermi level at half-filling.

\begin{figure*}[!hbt]
\centering
  \includegraphics[width=0.23\textwidth]{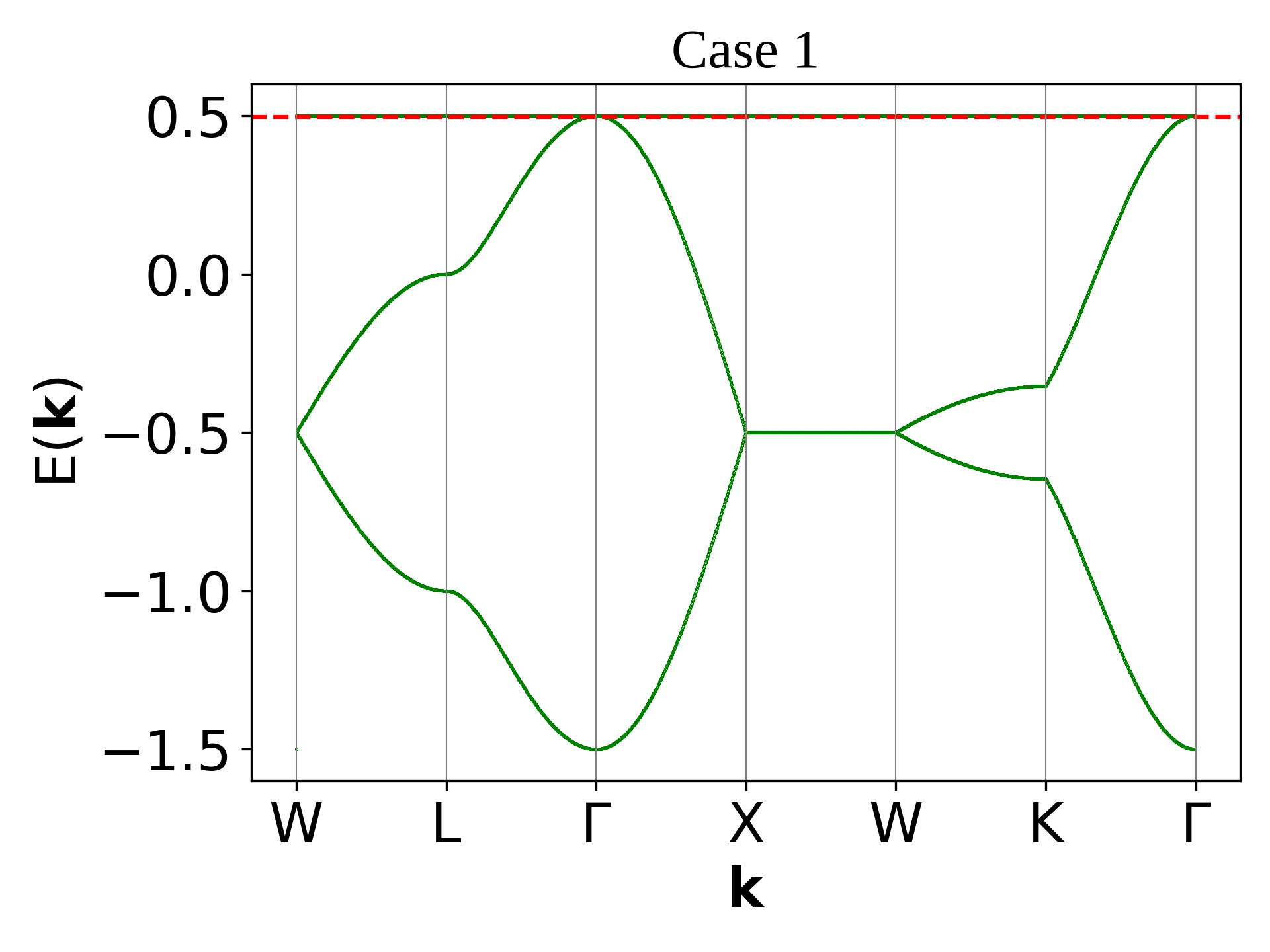}
    \hspace{0.1cm} 
     \includegraphics[width=0.23\textwidth]{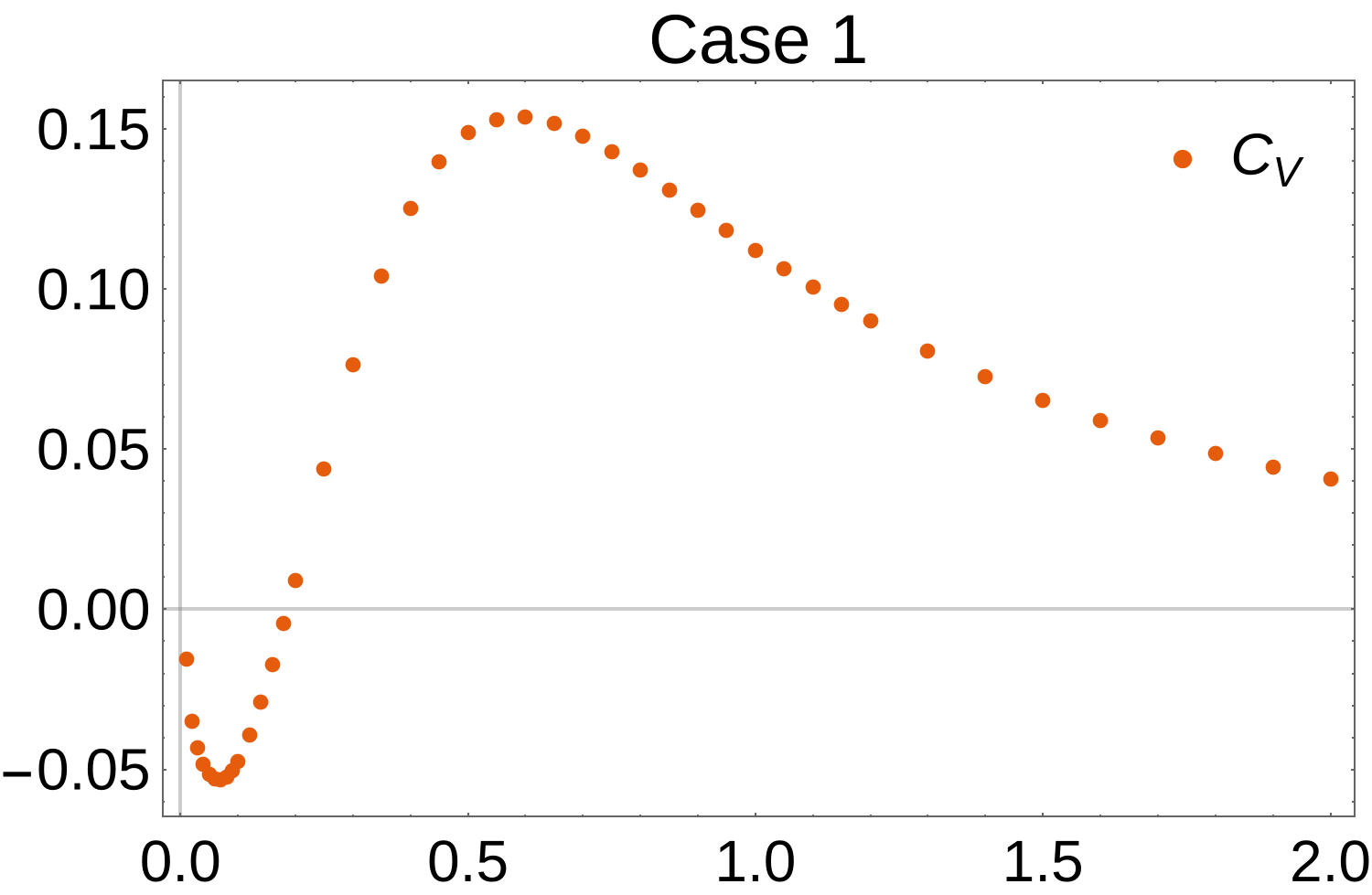}
     \hspace{0.1cm} 
     \includegraphics[width=0.23\textwidth]{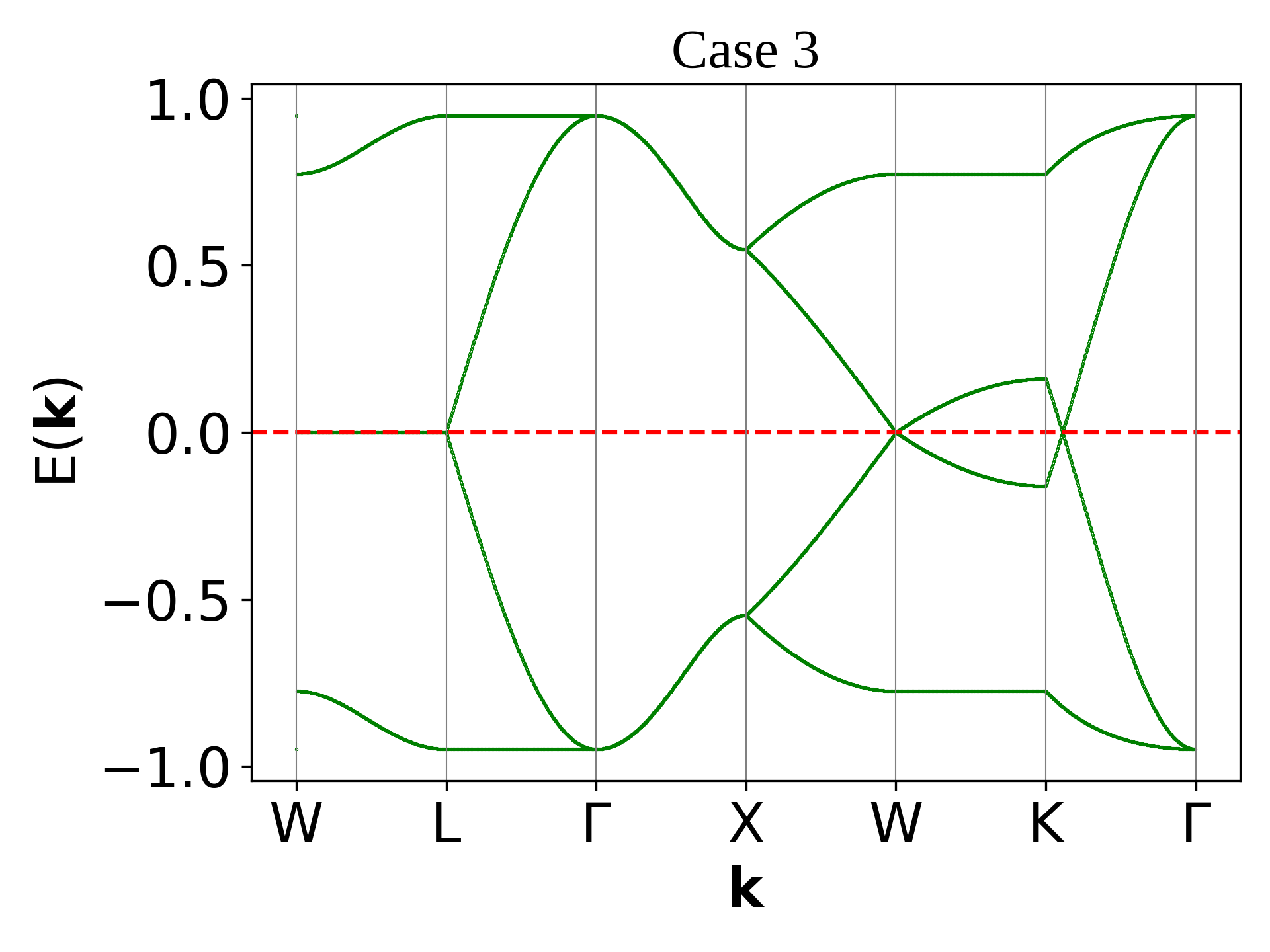}    
  \includegraphics[width=0.23\textwidth]{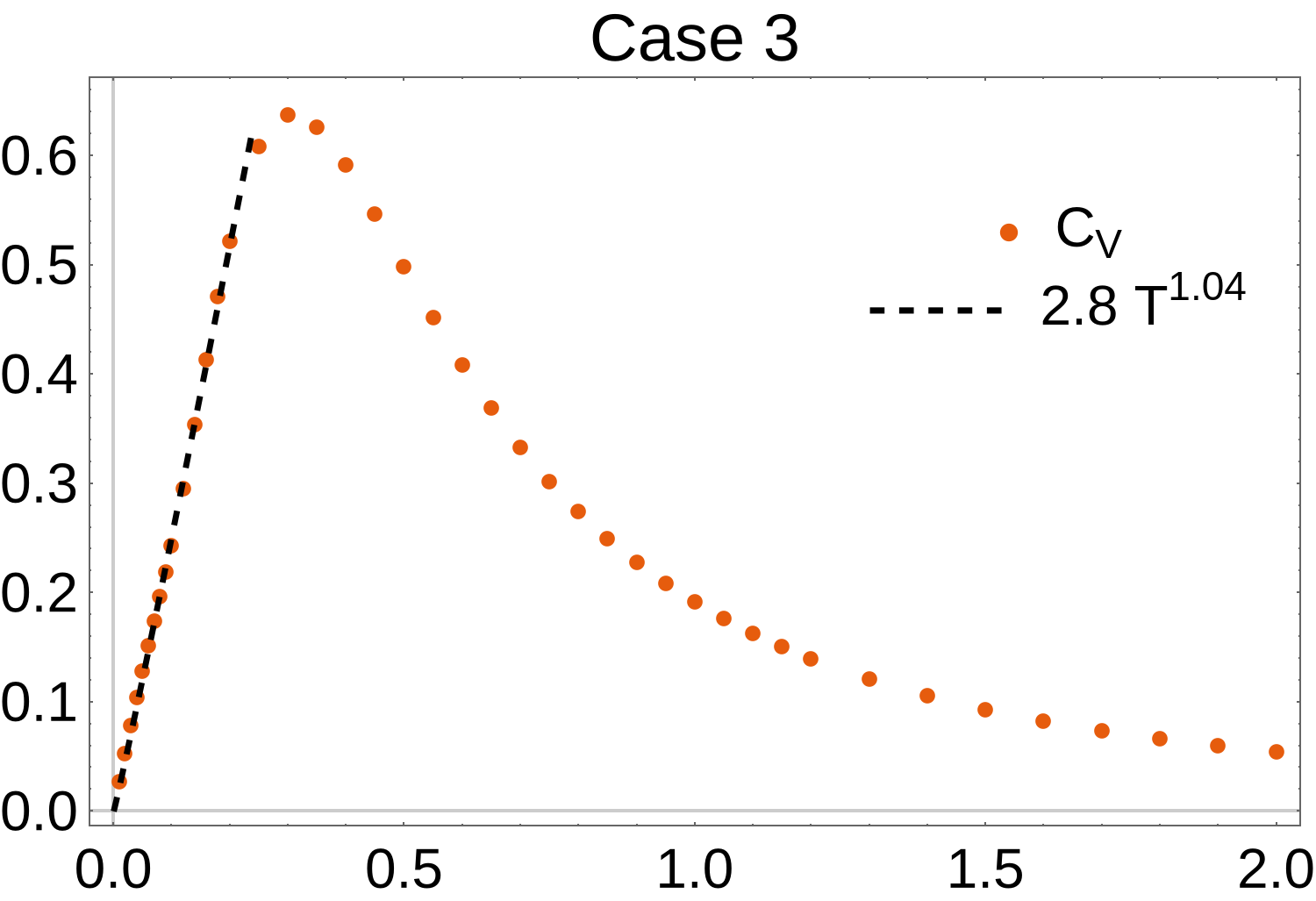}
    \hspace{0.1cm} 
     \includegraphics[width=0.23\textwidth]{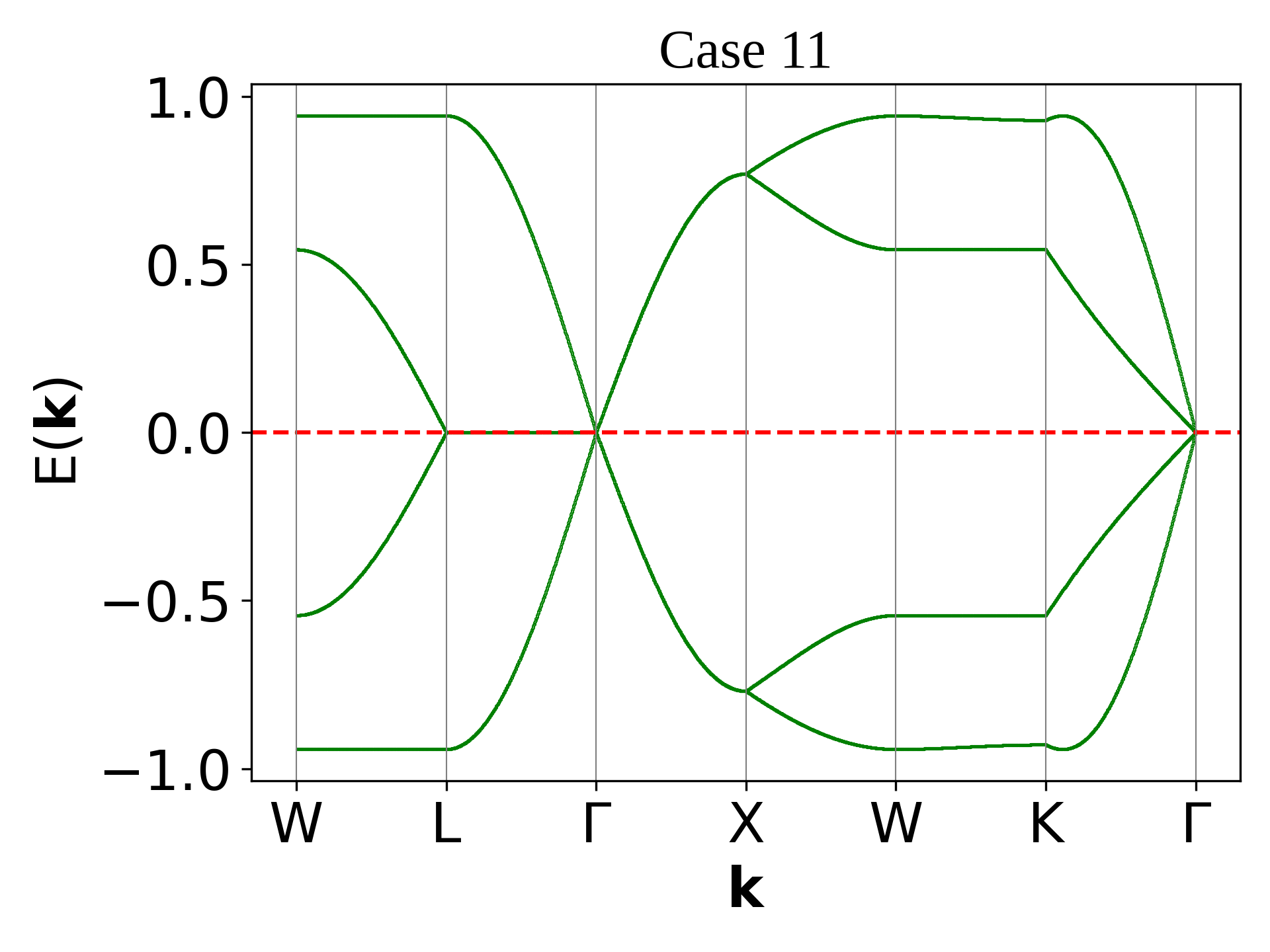}
     \hspace{0.1cm} 
     \includegraphics[width=0.23\textwidth]{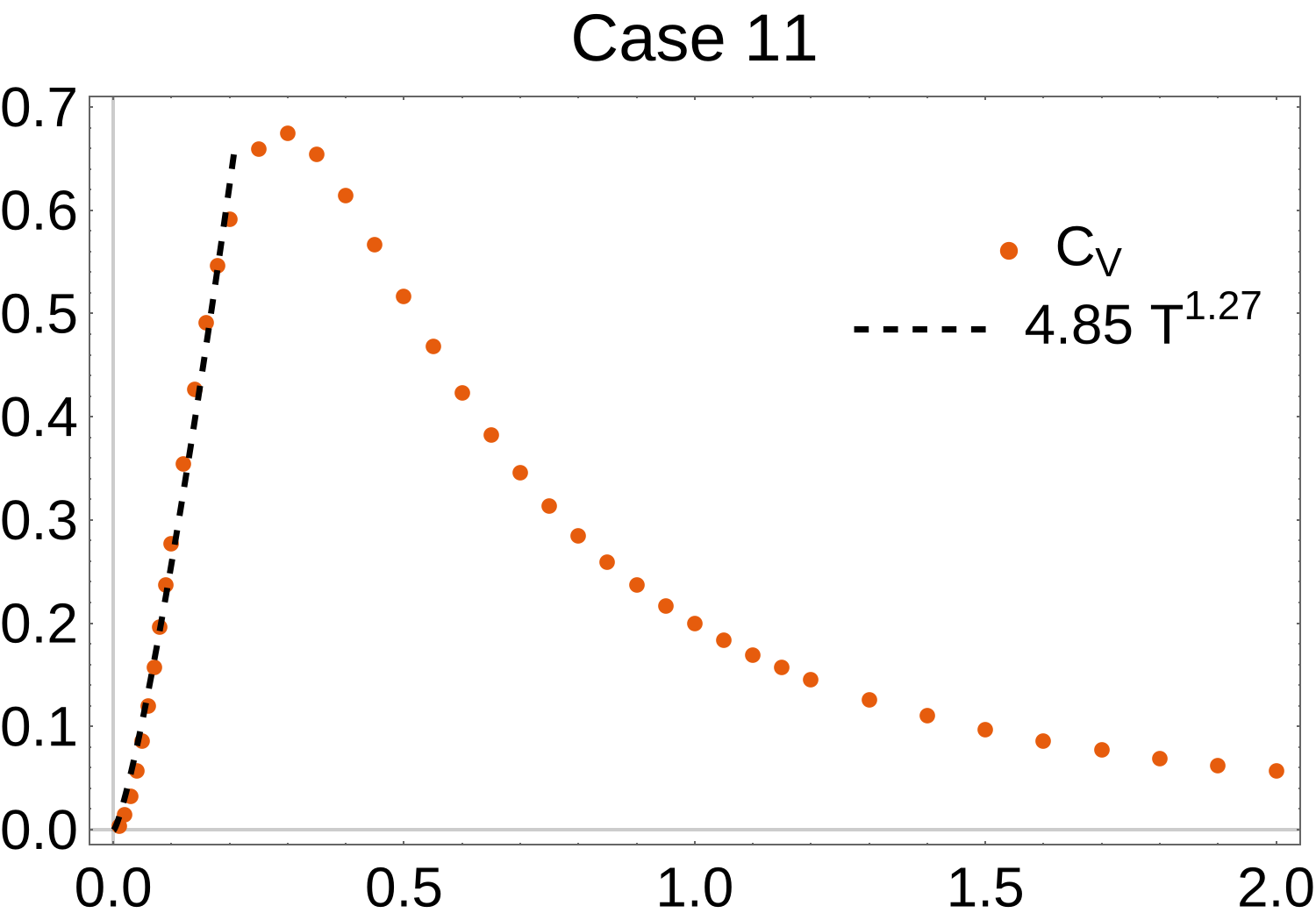}
     \includegraphics[width=0.23\textwidth]{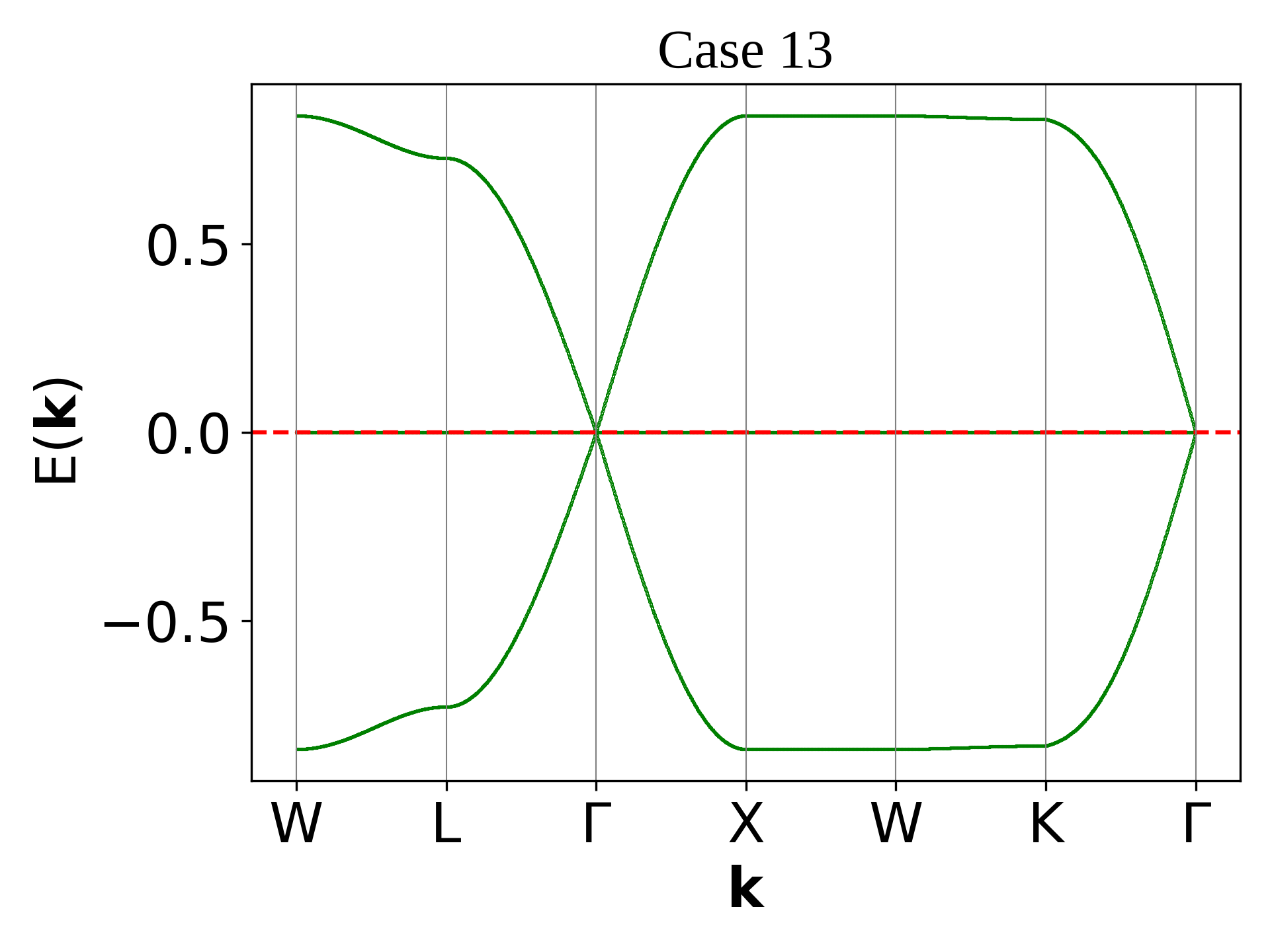}
     \hspace{0.1cm} 
     \includegraphics[width=0.23\textwidth]{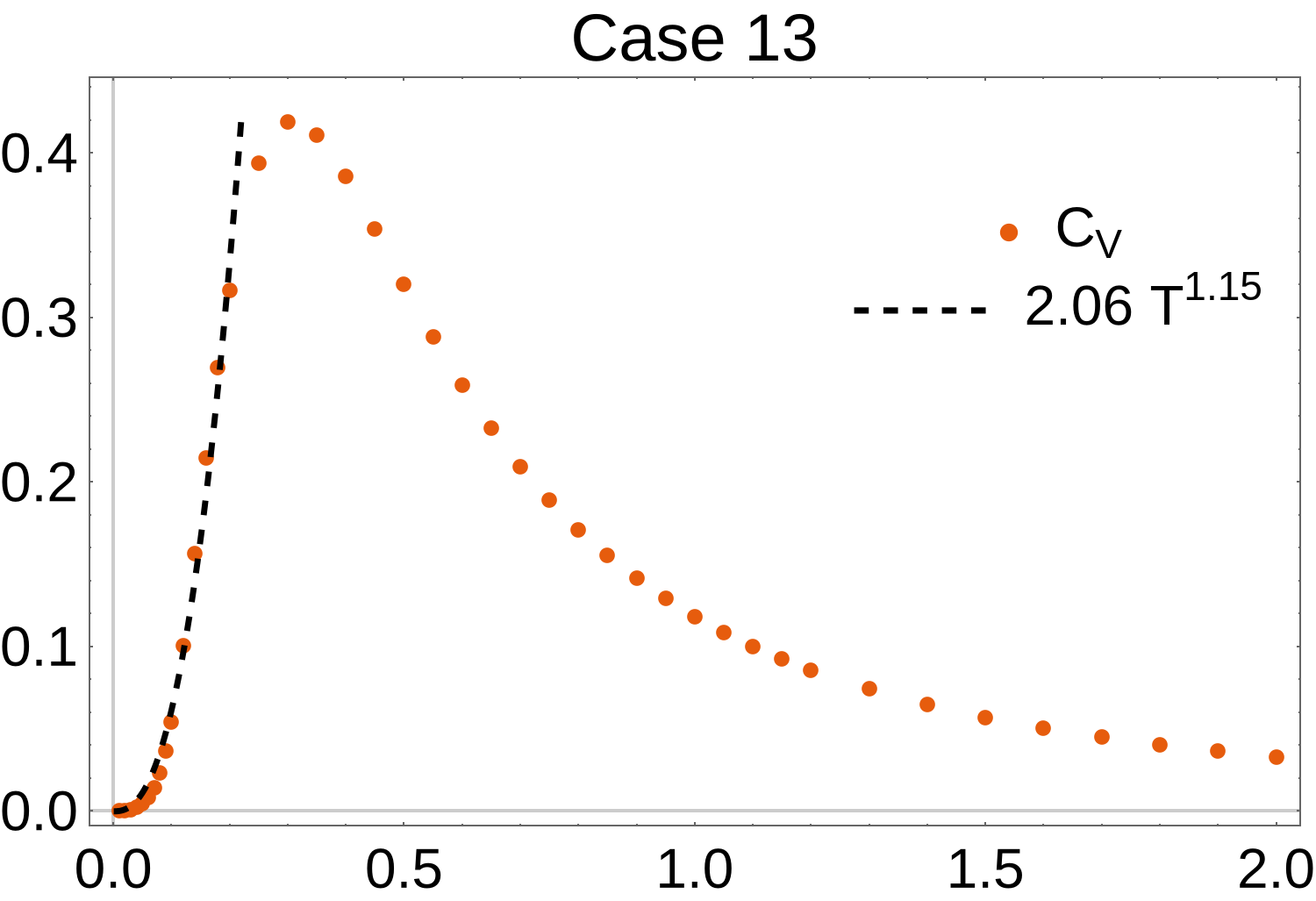}
     \caption{The energy dispersion and specific heat as a function of temperature are plotted side by side for PSG classes 1, 3, 11, and 13. The orange dotted line in energy dispersion plots denotes the fermi energy corresponding to the half-filling. The dotted black line in the specific heat plot is the fit.}
      \label{dispersion_spheat_fig}
\end{figure*}

In Fig. \ref{dispersion_spheat_fig} the temperature dependence of specific heat is given for the PSG classes 1, 3, 11, and 13. The PSG cases 3, 4, 7, 12, and 16 have nearly linear temperature dependence of specific heat. The temperature dependence is given in the table \ref{table_summary} for the rest of the PSG classes. The PSG classes 1 and 2 show an unusual upturn in specific heat, which is coming from the flat bands near the Fermi energy. \\\\
We also study the flux captured by the partons in each of the unit cells of the pyrochlore. We notice that, although all of these PSG classes are time-reversal symmetric, PSG classes 3, 4, 15, and 16 host a monopole flux state in each of the tetrahedra. Fig. \ref{odd_mf_figure} and \ref{even_mf_figure} captures the flux states for all the PSG classes. This is solely allowed because of the underlying 
dipolar-octupolar symmetry of the spin operators. For usual spin-1/2 fermionic partons, breaking of TR symmetry is necessary to host a monopole flux \cite{Burnell}.
\subsection{Structure factor}
\begin{figure}[!hbt]
  \includegraphics[width=0.23\textwidth]{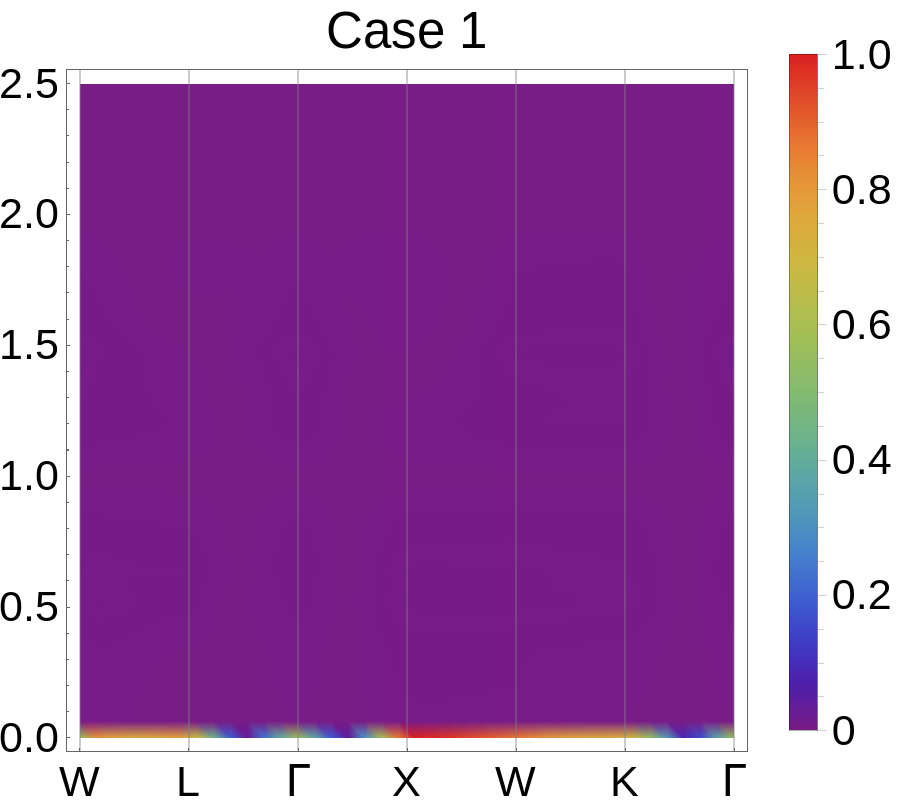}
    \hspace{0.1cm} 
     \includegraphics[width=0.23\textwidth]{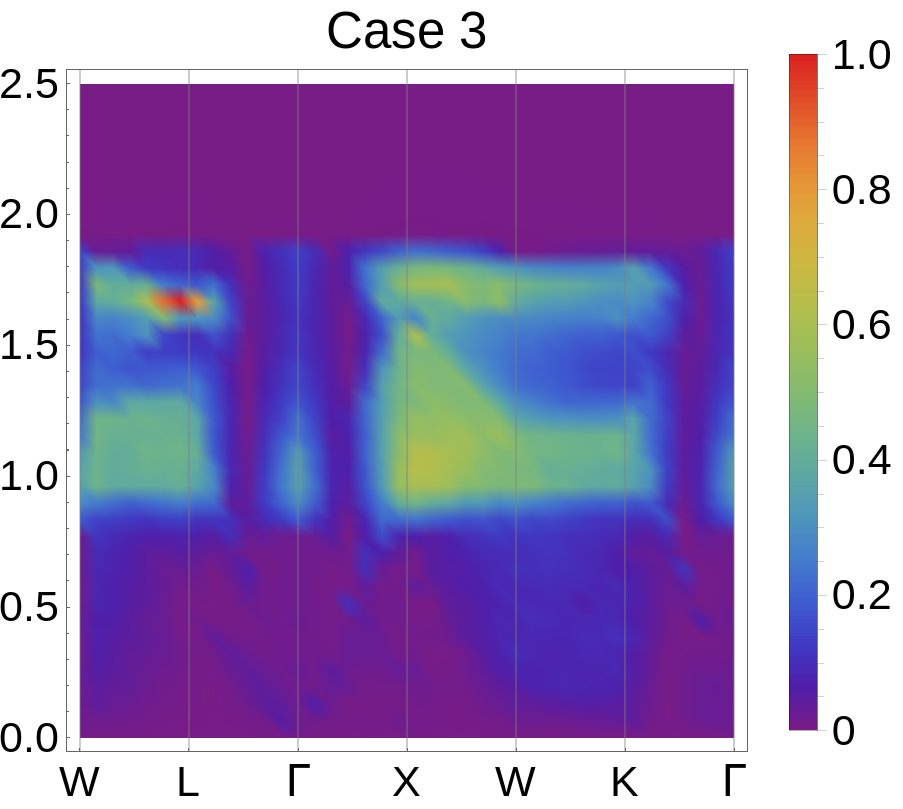}
     \hspace{0.1cm} 
  \includegraphics[width=0.23\textwidth]{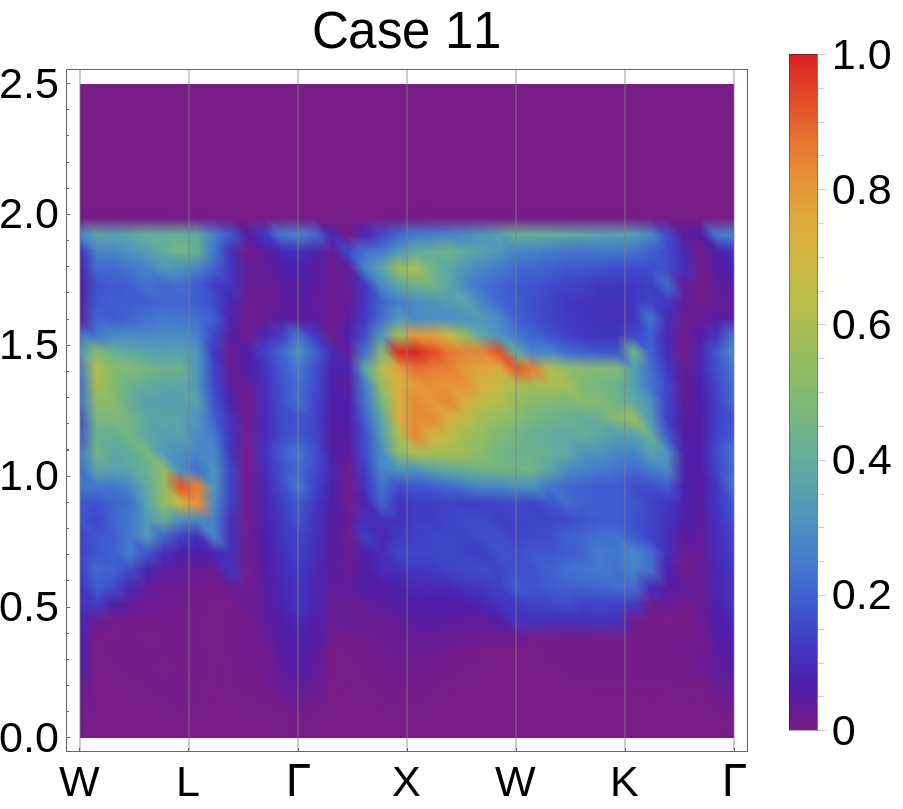}
    \hspace{0.1cm} 
     \includegraphics[width=0.23\textwidth]{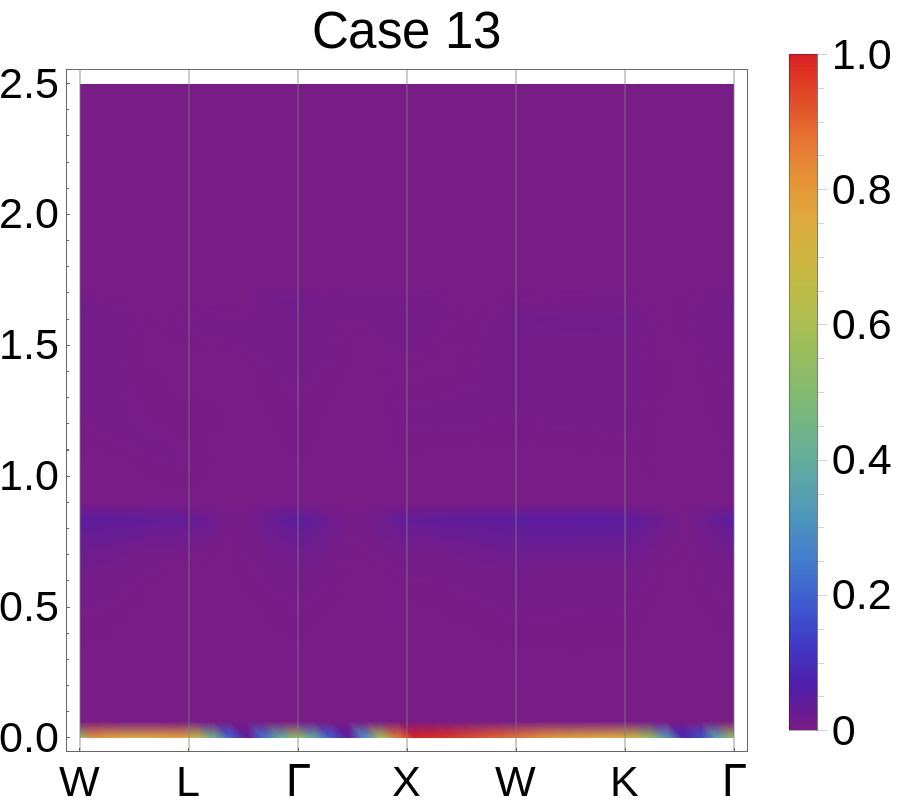}
     \caption{The dynamic structure factor for different PSG classes 1, 3, 11, and 13 are shown here.}
     \label{sf_figs}
\end{figure}
This section studies the dynamic spin-spin structure factor seen in neutron scattering experiments. The dynamic structure factor is defined as
\begin{equation}
\begin{split}
    &\Tilde{S}^{INS}(q,\omega)\\
    &= \sum_{a,b} \int_{-\infty}^{\infty} dt e^{i \omega t} \left(\delta_{ab}-\frac{q_a q_b}{\abs{\Vec{q}}^2}\right)\langle m^a(\Vec{q},t)m^b(-\Vec{q},0)\rangle\\
\end{split}
\end{equation}
The superscript INS to $\Tilde{S}$ means inelastic neutron scattering. And the summation indices a and b runs over $\{x,y,z\}$. And the $m^a(q,t)$ is defined as
\begin{equation}
    \begin{split}
    m^a(\Vec{q},t)& = \frac{1}{\sqrt{N_{u.c.}}}\sum_{\mathbf{r}_\mu}e^{i\Vec{q}.\mathbf{r}_\mu}\sum_{b}g_\mu^{ab}\Tilde{S}^b_{\mathbf{r}_\mu} \\
    & = \frac{1}{\sqrt{N_{u.c.}}}\sum_{\mathbf{r}_\mu}e^{i\Vec{q}.\mathbf{r}_\mu}~\hat{z}^a_\mu~\sum_{b}g^{\prime}_{b}S^b_{\mathbf{r}_\mu}
    \end{split}
\end{equation}
Where $\Tilde{S}^b_{\mathbf{r}_\mu}$ is the spin operator at $\mathbf{r}_\mu$ in global coordinate system. The matrix $g^{ab}_\mu$ is the coupling matrix, which tells how the magnetic moment of the neutron couples with the global spin operator. Also, $\mathbf{r}_\mu = \mathbf{r}+\Vec{d}_\mu$, where $\mathbf{r}$ is the translation vector of the pyrochlore lattice and the $\Vec{d}_\mu$ is the sublattice vector. The $\hat{z}^a_\mu$ is the local z-axis and $S^a_{\mathbf{r}_\mu}$ is the spin operator in the local coordinate system. And $g^{\prime}$ is given by $g^{\prime} =(g_x, g_y, g_z)$. Where,  $g_x, g_y, g_z$ are the coupling strength of neutron magnetic moment with the local $S^x, S^y, S^z$ operators.

We have plotted the dynamic structure factor in Fig. \ref{sf_figs} for PSG classes 1, 3, 11, and 13. Case 1 and Case 13 shows excitations at very low energy at all momenta. It is due to the presence of two degenerate flat bands right at the Fermi level. Moreover, in Fig. \ref{dispersion_spheat_fig} the gapless excitation for Case 3 is near W, K, and for Case 11 it is near $\Gamma$ point, respectively. This feature is also captured in the low-energy structure factor near the respective band touching points in Fig \ref{sf_figs}.

\section{Discussion}
\label{summary}
We conclude this article with a brief review of other related works in the literature, followed by a summary of our most important results and a discussion of several possible directions for future work. 

\subsection{Relation to other works}
Our results can be contrasted with the U(1) and $Z_2$ fermionic PSG\cite{Liu_fermion} and $Z_2$ bosonic\cite{Liu_boson} classification of spin liquids for spin-1/2 particles on pyrochlore lattice. In the fermionic PSG, 18 $U(1)$ spin liquids and 48 $Z_2$ spin liquids preserve the spacegroup symmetries of pyrochlore lattice and time-reversal symmetry. All the fully symmetric solutions bind either a $0$ or $\pi$ flux; the time-reversal symmetry broken solutions bind a flux
$\pi/2$. Many of these spinon Hamiltonians possess gapless nodal lines along the four
equivalent (111) directions of the Brillouin zone, dubbed “nodal star” spin liquid. The bare spinon contribution to low temperatures specific heat for these "nodal star" spin liquids was shown to be $C~ \sim T^{3/2}$. 
In contrast, for bosonic $Z_2$ spinons, there are 16 different QSL phases, of these six are with $0$ flux with bare spinon contribution to specific heat as a power law with $C~ \sim T^x$ where $x=1,3/2,2,3$.\\

Another recent study discusses the complete PSG classification with bosonic spinons for spin models on  DO pyrochlore\cite{DO_Kim}. At the nearest-neighbor coupling, there are the same 16 $Z_2$ QSLs and 4 $U(1)$ QSLs. The $U(1)$ QSLs are predicted to have $C~ \sim T^{3}$ temperature dependence of heat capacity; the $Z_2$ case is expected to have a thermally activated exponential form.  More recently\cite{Chern2023} a XYZ Hamiltonian of dipolar-octupolar pyrochlore magnets is studied with the pseudofermion functional renormalization group (PFFRG), which suggests a stable phase with fermionic spinons. 
\subsection{Summery of results }
In this work, we studied the U(1) fermionic spin liquid with dipolar-octupolar symmetry on a pyrochlore lattice. Using the PSG formalism, we calculated the symmetry allowed nearest neighbor mean-field processes for each of the 16 time-reversal symmetric PSG classes and solved them variationally. We found that no mean-field solutions exist for four of the PSG classes. For six of the PSG classes, the unit cell is that of a regular pyrochlore lattice, and for the other six, the unit cell is four times larger than the pyrochlore lattice unit cell. 

Importantly, out of the 12 PSG classes with finite mean-field solutions, four PSG classes bind a monopole flux state in each pyrochlore tetrahedra. In the case of effective spin 1/2 moments, the monopole flux state breaks the TR symmetry and hence is not allowed under time-reversal symmetric PSG. However, it is allowed in a time-reversal symmetric PSG because of the dipolar-octupolar symmetry of the underlying spin states.  

Moreover, we also studied the temperature dependence of specific heat and the dynamical spin structure factor. For cases 3, 4, 7, 12, and 16, we found the low-temperature specific heat has a linear temperature dependence. Cases 1 and 13 show gapless excitation in structure factor at all momenta.

\subsection{Conclusion and future directions}
\label{sec_conclusion}
Our study points out the features of a dipolar-octupolar spin liquid using U(1) fermionic parton ansatz that can be captured in specific heat measurements and neutron scattering experiments. We propose several threads of study that will be of immediate relevance to our work. Our calculation involves nearest neighbor mean-field theory; adding at least one next-neighbor is an immediate next step to understand further possible phases in  U(1) fermionic parton ansatz. Furthermore, the stability of our mean-field solutions needs to be understood in the presence of beyond mean-field fluctuations, such as implementing the filling constraint (eqn. \ref{filling}) at each site using Gutzwiller projection. Another important direction will be to compare the energies of the bosonic and fermionic spinon ansatz in different exchange parameter values of the XYZ model to predict possible, stable mean-field solutions at different regimes of the phase diagrams. In any solid-state experiments, the effects of phonon coupling are ubiquitous. Hence, it will be worthwhile to understand the stability of these spinon solutions in the presence of phonon-spinon coupling. 
\begin{acknowledgments}
S.~S. acknowledges fruitful discussions with Subhro Bhattacharjee. S.~S. acknowledges support from the Science and Engineering Research Board (Department of Science and Technology) Govt. of India, under grant no. SRG/2020/001525 and an internal start-up grant from the Indian Institute of Science Education and Research, Tirupati. 
\end{acknowledgments}

\appendix
 \section{Local and global axes of spin quantization}\label{local-global}
The global coordinate axes are given by $\hat{x}=(1,0,0), \hat{y}=(0,1,0), \hat{z}=(0,0,1)$. The following table lists the Local coordinate system at each sublattice $\mu$. The transformation for the spin operators from local ($S_{\mathbf{r}_\mu}$) to global($\tilde{S}_{\mathbf{r}_\mu}$) spin quantization axis is in \ref{local_table}.
\begingroup
\setlength{\tabcolsep}{3 pt}
\renewcommand{\arraystretch}{1.5} 
\begin{table}
\begin{center}
    \begin{tabular}{c c c c c}
         \hline
         $\mu$ & 0 & 1 & 2 & 3 \\
         \hline
         $\hat{x}_{\mu}$ & $\frac{-1}{\sqrt{6}}(-2,1,1)$ & $\frac{-1}{\sqrt{6}}(1,2,-1)$ & $\frac{-1}{\sqrt{6}}(-1,1,2)$ & $\frac{1}{\sqrt{6}}(1,1,2)$ \\
         $\hat{y}_{\mu}$ & $\frac{1}{\sqrt{2}}(0,-1,1)$ & $\frac{1}{\sqrt{2}}(1,0,1)$ & $\frac{1}{\sqrt{2}}(1,1,0)$ & $\frac{1}{\sqrt{2}}(1,-1,0)$ \\
         $\hat{z}_{\mu}$ & $\frac{1}{\sqrt{3}}(1,1,1)$ & $\frac{1}{\sqrt{3}}(-1,1,1)$ & $\frac{1}{\sqrt{3}}(1,-1,1)$ & $\frac{1}{\sqrt{3}}(1,1,-1)$ \\
         \hline
    \end{tabular}
    \caption{This table lists the local basis vectors for each of the sublattices.}
    \label{local_table}
\end{center}
\end{table}
\endgroup
]
\begin{tabular} {c c}
    $
        R_0 = \begin{pmatrix}
            \sqrt{\frac{2}{3}} & -\frac{1}{\sqrt{6}} & -\frac{1}{\sqrt{6}} \\
            0 & -\frac{1}{\sqrt{2}} & \frac{1}{\sqrt{2}} \\
            -\frac{1}{\sqrt{3}} & -\frac{1}{\sqrt{3}} & -\frac{1}{\sqrt{3}}\\
        \end{pmatrix}
    $
        &
    $
        R_1 = \begin{pmatrix}
            -\frac{1}{\sqrt{6}} & -\sqrt{\frac{2}{3}} & \frac{1}{\sqrt{6}} \\
             \frac{1}{\sqrt{2}} & 0 & \frac{1}{\sqrt{2}} \\
            -\frac{1}{\sqrt{3}} & \frac{1}{\sqrt{3}} & \frac{1}{\sqrt{3}}\\
        \end{pmatrix}
    $
        \\
    $
        R_2 = \begin{pmatrix}
            \frac{1}{\sqrt{6}} & -\frac{1}{\sqrt{6}} & -\sqrt{\frac{2}{3JIJ}}  \\
             \frac{1}{\sqrt{2}} & \frac{1}{\sqrt{2}} & 0 \\
            \frac{1}{\sqrt{3}} & -\frac{1}{\sqrt{3}} & \frac{1}{\sqrt{3}}\\
        \end{pmatrix}
     $
                & 
    $
                R_3 = \begin{pmatrix}
            \frac{1}{\sqrt{6}} & \frac{1}{\sqrt{6}} & \sqrt{\frac{2}{3}}  \\
             \frac{1}{\sqrt{2}} & -\frac{1}{\sqrt{2}} & 0 \\
            \frac{1}{\sqrt{3}} & \frac{1}{\sqrt{3}} & -\frac{1}{\sqrt{3}}\\
        \end{pmatrix} 
    $
        \\
    \end{tabular}
The Hamiltonian in the local frame with an external magnetic field $\Vec{h}$ is given by:
\begin{eqnarray*}
H_{L} &=& \sum_{\langle \mathbf{r}_\mu,\mathbf{r}^{\prime}_\nu \rangle} J_{xx}S_{\mathbf{r}_\mu}^xS_{\mathbf{r}^{\prime}_\nu}^x+J_{yy}S_{\mathbf{r}_\mu}^yS_{\mathbf{r}^{\prime}_\nu}^y+J_{zz}S_{\mathbf{r}_\mu}^zS_{\mathbf{r}^{\prime}_\nu}^z \\
&+&J_{xz}(S_{\mathbf{r}_\mu}^xS_{\mathbf{r}^{\prime}_\nu}^z+S_{\mathbf{r}_\mu}^zS_{\mathbf{r}^{\prime}_\nu}^x) \\
&-& \sum_{\mathbf{r}_\mu}(\hat{z}_\mu.\Vec{h})(g_xS^x_{\mathbf{r}_\mu}+g_zS^z_{\mathbf{r}_\mu}).
\end{eqnarray*}
In the global frame the Hamiltonian is given by,
\begin{equation*}
H_G=\sum_{\langle \mathbf{r}_\mu,\mathbf{r}^{\prime}_\nu \rangle}\sum_{ab}J_{\mu\nu}^{ab}\Tilde{S}_{\mathbf{r}_\mu}\Tilde{S}_{\prime{\mathbf{r}}_\nu}-\sum_{\mathbf{r}_\mu}\sum_{ab}h^{a}g_{\mu}^{ab}\Tilde{S}_{\mathbf{r}_\mu},
\end{equation*}
where $a\epsilon\{x,y,z\}$ and $\mu,\nu\epsilon\{0,1,2,3\}$. The coupling matrices $J_{\mu \nu}$ are
\begin{center}
    \begin{tabular}{c c}
        $
             J_{01}=\begin{pmatrix}
                 J_1 & J_2 & -J_1 \\
                 J_3 & -J_1 & J_4 \\
                 -J_4 & -J_1 & -J_3 \\
             \end{pmatrix}
        $
         & 
         $
             J_{02}=\begin{pmatrix}
                 -J_1 & J_1 & J_2 \\
                 J_4 & J_3 & -J_1 \\
                 -J_3 & -J_4 & -J_1 \\
             \end{pmatrix}
         $
         \\
        
         $
             J_{03}=\begin{pmatrix}
                 -J_1 & -J_1 & -J_2 \\
                 J_4 & -J_3 & J_1 \\
                 -J_3 & J_4 & J_1 \\
             \end{pmatrix}
         $
         & 
        $
             J_{12}=\begin{pmatrix}
                 -J_3 & -J_4 & -J_1 \\
                 J_1 & -J_1 & -J_2 \\
                 -J_3 & -J_4 & J_1 \\
             \end{pmatrix}
         $
         \\
         
        $
             J_{13}=\begin{pmatrix}
                 -J_3 & J_4 & J_1 \\
                 J_1 & J_1 & J_2 \\
                 -J_4 & J_3& -J_1 \\
             \end{pmatrix}
         $
         & 
        $
             J_{23}=\begin{pmatrix}
                 -J_4 & J_4 & -J_1 \\
                 -J_3 & J_4 & J_1 \\
                 J_1 & J_1 & J_2 \\
             \end{pmatrix},
        $
         \\
    \end{tabular}
\end{center}
where 
\begin{equation*}
    \begin{split}
        & J_1 = \frac{1}{6}(-2J_{xx}+2J_{zz}-\sqrt{2}J_{xz})\\
        & J_2 = \frac{1}{3}(-2J_{xx}-J_{zz}+2\sqrt{2}J_{xz})\\
        & J_3 = \frac{1}{6}(J_{xx}-3J_{yy}+2J_{zz}+2\sqrt{2}J_{xz})\\
        & J_4 = \frac{1}{6}(-J_{xx}-3J_{yy}-2J_{zz}-2\sqrt{2}J_{xz}).\\
    \end{split}
\end{equation*}
And
\begin{center}
    \begin{tabular}{c c}
        $
             g_{0}=\begin{pmatrix}
                 -g_{-} & g_+ & g_+ \\
                 -g_{-} & g_+ & g_+ \\
                 -g_{-} & g_+ & g_+ \\
             \end{pmatrix}
         $
         &  
         $
             g_{1}=\begin{pmatrix}
                 g_{+} & g_- & -g_+ \\
                 -g_{+} & -g_- & g_+ \\
                 -g_{+} & -g_- & g_+ \\
             \end{pmatrix}
         $\\
          
         $
             g_{2}=\begin{pmatrix}
                 g_{+} & -g_+ & -g_- \\
                 -g_{+} & g_+ & g_- \\
                 g_{+} & -g_+ & -g_- \\
             \end{pmatrix}
        $
         &  
        $
             g_{1}=\begin{pmatrix}
                 g_{+} & g_+ & g_- \\
                 g_{+} & g_+ & g_- \\
                 -g_{+} & -g_+ & -g_- \\
             \end{pmatrix},
         $\\
    \end{tabular}
\end{center}
where $g_+=\frac{1}{3}(\frac{g_x}{\sqrt{2}}+g_z)$,$g_-=\frac{1}{3}({\sqrt{2}}g_x-g_z)$.

\section{dipolar-octupolar Symmetry}\label{DO_Symmetry}
The pyrochlore lattice has $D_{3d}$ {site} symmetry, composed of a $C_3$ rotation, three $C_2$ rotations, inversion, and various combinations. The frustration in rare-earth pyrochlore comes from its tetrahedral geometry and highly anisotropic spin-orbit coupling interactions in the system. In this system, we can use the total angular momentum quantum number J to label the ground-state manifold of the single ion. However, the ground state manifold degeneracy is lifted by the surrounding crystal electric field (CFT) and the lowest-lying states form a doublet($\ket{\pm}$). The pseudo spin operators are defined as, 
$ S^z=\frac{\ket{+}\bra{+}-\ket{-}\bra{-}}{2}$ and $ S^{\pm}=\ket{\pm}\bra{\mp}$. The irreducible representation of $D_{3d}$ group further classifies these doublets. We focus on dipolar-octupolar doublet, which transforms as $\Gamma_5\oplus\Gamma_6$.

The pseudo spin operators transform under the space group symmetry operations. The symmetry operation mixes the sublattices, which in turn requires changing the local coordinate system. Therefore, the pseudo spin transformation includes a unitary operation accounting for the symmetry operation and another unitary operation that keeps track of the change of the local coordinate system due to sublattice mixing. 

The $\bar{C}_{6}$ and $S$ operation on $\ket{\pm}$ is achieved with
\begin{equation*}
    R_{\bar{C}_{6},\mu} = e^{-i\frac{2\pi}{3}\hat{n}_{\bar{C}_{6},\mu}.\Vec{J}}, ~~ R_{S,\mu} = e^{-i\pi\hat{n}_{S,\mu}.\Vec{J}}.
\end{equation*}
Where, the $\hat{n}_{\mathcal{O},\mu}$ denotes the axes of rotation in local frame at $\mu$th sublattice. In global frame, $\hat{n}_{\bar{C}_{6}}=\frac{1}{\sqrt{3}}(1,1,1)$ and $\hat{n}_{S}=\frac{1}{\sqrt{2}}(1,1,0)$. 

The change in the local coordinate system is accounted by 
\begin{equation*}
    R_{\mu \rightarrow \nu} = e^{-i\theta_\alpha J_z} e^{-i\theta_\beta J_y} e^{-i\theta_\gamma J_z}.
\end{equation*}
Where $\theta_\alpha, \theta_\beta, \theta_\gamma$ are the euler angles that takes from $\mu$th to $\nu$th local-coordinate system. 

The full transformation is given by
\begin{equation*}
    U_{\mathcal{O},\mu} = \mathcal{P}R_{\mu \rightarrow \mathcal{O}(\mu)} R_{\bar{C}_{6},\mu} \mathcal{P}
\end{equation*}
Where $\mathcal{P}$ is the projection operator which project into $\ket{\pm}$. The explicit calculation of the above expression leads to the following form of the $U_{\mathcal{O}}$ operators for DO spin:
\begin{equation*} U_{T_i}=\mathbf{1}_{2\cross2},~~U_{\bar{C}_{6}}=\mathbf{1}_{2\cross2},~~U_S=-i\sigma_y,~~U_{\mathcal{T}} = i\sigma_y.
\end{equation*}
Note the $U_{\mathcal{O}}$ operators are all sublattice independent which is a characteristic feature of DO spins.

\section{Generic mean-field decomposition}\label{generic mf decomposition}
In this section we will give an outline for writing the spin Hamiltonian in terms of different fermionic mean-field channels. We start with the spin Hamiltonian,
\begin{equation}\label{eq1}
    H = \sum_{\langle i,j \rangle}J_{ij}\mathbf{S}_i.\mathbf{S}_j
\end{equation}
Doing the parton construction,
\begin{equation}\label{eq2}
\mathbf{S}_i=\frac{1}{2}f_{i\alpha}^{\dagger}\mathbf{\sigma}_{\alpha\beta}f_{i\beta}
\end{equation}
The spin Hamiltonian changes to a Hamiltonian with four-fermi terms. The goal here is to write this quartic fermionic Hamiltonian in terms of all possible symmetry-preserving quadratic mean-field channels given in Eq \ref{mf_channels}. We start by expressing spin Hamiltonian ( Eq. {\ref{eq1}}) in terms of the fermionic operators in normal order:
\begin{equation}\label{eq5}
    \begin{split}
       & J_x S_i^x S_j^x = -\frac{J_x}{4}(f_{i\downarrow}^\dagger f_{j\downarrow}^\dagger f_{i\uparrow}f_{j\uparrow}+f_{i\downarrow}^\dagger f_{j\uparrow}^\dagger f_{i\uparrow}f_{j\downarrow}\\
       &+f_{i\uparrow}^\dagger f_{j\downarrow}^\dagger f_{i\downarrow}f_{j\uparrow}+f_{i\uparrow}^\dagger f_{j\uparrow}^\dagger f_{i\downarrow}f_{j\downarrow}) \\
       & J_y S_i^y S_j^y = \frac{J_y}{4}(f_{i\downarrow}^\dagger f_{j\downarrow}^\dagger f_{i\uparrow}f_{j\uparrow}-f_{i\downarrow}^\dagger f_{j\uparrow}^\dagger f_{i\uparrow}f_{j\downarrow}\\
       &-f_{i\uparrow}^\dagger f_{j\downarrow}^\dagger f_{i\downarrow}f_{j\uparrow}+f_{i\uparrow}^\dagger f_{j\uparrow}^\dagger f_{i\downarrow}f_{j\downarrow}) \\
       & J_z S_i^z S_j^z = \frac{J_z}{4}(-f_{i\uparrow}^\dagger f_{j\uparrow}^\dagger f_{i\uparrow}f_{j\uparrow}+f_{i\uparrow}^\dagger f_{j\downarrow}^\dagger f_{i\uparrow}f_{j\downarrow}\\
       &+f_{i\downarrow}^\dagger f_{j\uparrow}^\dagger f_{i\downarrow}f_{j\uparrow}-f_{i\downarrow}^\dagger f_{j\downarrow}^\dagger f_{i\downarrow}f_{j\downarrow}). \\
    \end{split}
\end{equation}
Consider the following generic expression where operators are the mean-field channels and write them in the same order in which the spin operators are written in \ref{eq5}. 
\begin{equation}\label{eq6}
    \begin{split}
        & A_1\hat{\chi}_{ij}^\dagger\hat{\chi}_{ij}+A_2\hat{\Delta}_{ij}^\dagger\hat{\Delta}_{ij}+A_3{\hat{E}_{ij}}^{x\dagger }\hat{E}_{ij}^x+A_4{\hat{E}_{ij}}^{y\dagger} \hat{E}_{ij}^y\\
        &+A_5{\hat{E}_{ij}}^{z\dagger} \hat{E}_{ij}^z+A_6{\hat{D}_{ij}}^{x\dagger} \hat{D}_{ij}^x +A_7{\hat{D}_{ij}}^{y\dagger} \hat{D}_{ij}^y\\
        &+A_8{\hat{D}_{ij}}^{z\dagger} \hat{D}_{ij}^z+A_9\hat{n}_i\hat{n}_j+A_{10}\hat{n}_j \\
        & = \hat{n}_j(A_1+A_3+A_4+A_5+A_{10})\\
        & +(f_{i\uparrow}^\dagger f_{j\uparrow}^\dagger f_{i\uparrow}f_{j\uparrow}+f_{i\downarrow}^\dagger f_{j\downarrow}^\dagger f_{i\downarrow}f_{j\downarrow})(A_1+A_5-A_6-A_7-A_9)\\
        &+(f_{i\downarrow}^\dagger f_{j\uparrow}^\dagger f_{i\uparrow}f_{j\downarrow}+f_{i\uparrow}^\dagger f_{j\downarrow}^\dagger f_{i\downarrow}f_{j\uparrow})(A_1+A_2-A_5-A_8)\\
        & +(f_{i\downarrow}^\dagger f_{j\uparrow}^\dagger f_{i\downarrow}f_{j\uparrow}+f_{i\uparrow}^\dagger f_{j\downarrow}^\dagger f_{i\uparrow}f_{j\downarrow})(-A_2+A_3+A_4-A_8-A_9)\\
        & +(f_{i\uparrow}^\dagger f_{j\uparrow}^\dagger f_{i\downarrow}f_{j\downarrow}+f_{i\downarrow}^\dagger f_{j\downarrow}^\dagger f_{i\uparrow}f_{j\uparrow})(A_3-A_4+A_6-A_7)\\
    \end{split}
\end{equation}
\textbf{AFM coupling i.e. $\mathbf{J_x, J_y, J_z > 0}$}

In order to decompose the spin-spin interaction terms into the fermionic mean-field operator, the coefficients of the four-fermi terms in Eq. (\ref{eq5}) are compared with the coefficients of four-fermi terms in Eq. (\ref{eq6}). 
\begin{equation}\label{eq7}
\begin{split}
    & A_{10}^a=-(A_1^a+A_3^a+A_4^a+A_5^a)\\
    & A_9^a=-\frac{1}{4}-A_1^a-2A_2^a+A_3^a+A_4^a+A_5^a\\
    & A_8^a = \frac{1}{4}(1-\delta_{a,z})+A_1^a+A_2^a-A_5^a\\
    & A_7^a = \frac{1}{4}(1-\delta_{a,y})+A_1^a+A_2^a-A_4^a\\
    & A_6^a = \frac{1}{4}(1-\delta_{a,x})+A_1^a+A_2^a-A_3^a\\
\end{split}
\end{equation}
Where the superscript $a \in \{x,y,z\}$ to the coefficients A are put to distinguish the coefficients coming from $S_i^x S_j^x$, $S_i^y S_j^y$, $S_i^z S_j^z$.
Using above we can write
\begin{equation}
    \begin{split}\label{eq10}
    & ~~~~J_x S_i^x S_j^x+J_y S_i^y S_j^y +J_z S_i^z S_j^z \\
    & = \chi_{ij}^\dagger\chi_{ij}(J_x A_1^x +J_y A_1^y +J_z A_1^z)\\
    &+\Delta_{ij}^\dagger\Delta_{ij}(J_x A_2^x +J_y A_2^y +J_z A_2^z)\\
    & +{E_{ij}^x}^\dagger E_{ij}^x(J_x A_3^x +J_y A_3^y +J_z A_3^z)\\
    &+{E_{ij}^y}^\dagger E_{ij}^y(J_x A_4^x +J_y A_4^y +J_z A_4^z)\\
    &+{E_{ij}^z}^\dagger E_{ij}^z(J_x A_5^x +J_y A_5^y +J_z A_5^z) \\
    &+{D_{ij}^x}^\dagger D_{ij}^x[J_x(A_1^x+A_2^x-A_3^x)\\
    &+J_y(\frac{1}{4}+A_1^y+A_2^y-A_3^y)+J_z(\frac{1}{4}+A_1^z+A_2^z-A_3^z)]\\
    &+{D_{ij}^y}^\dagger D_{ij}^y[J_x(\frac{1}{4}+A_1^x+A_2^x-A_4^x)\\
    &+J_y(A_1^y+A_2^y-A_4^y)+J_z(\frac{1}{4}+A_1^z+A_2^z-A_4^z)]\\
    &+{D_{ij}^z}^\dagger D_{ij}^z[J_x(\frac{1}{4}+A_1^x+A_2^x-A_5^x)\\
    &+J_y(\frac{1}{4}+A_1^y+A_2^y-A_5^y)+J_z(A_1^z+A_2^z-A_5^z)]\\
    &+n_in_j[J_x(-\frac{1}{4}-A_1^x-2A_2^x+A_3^x+A_4^x+A_5^x)\\
    &+J_y(-\frac{1}{4}-A_1^y-2A_2^y+A_3^y+A_4^y+A_5^y)\\
    &+J_z(-\frac{1}{4}-A_1^z-2A_2^z+A_3^z+A_4^z+A_5^z)]\\
    &-n_j[J_x(A_1^x+A_3^x+A_4^x+A_5^x)\\
    &+J_y(A_1^y+A_3^y+A_4^y+A_5^y)\\
    &+J_z(A_1^z+A_3^z+A_4^z+A_5^z)] \\
    \end{split}
\end{equation}
Equation (\ref{eq7}) is the most general mean-field decomposition along different mean-field channels.

In the $J_x=J_y=J_z$ limit, we have a huge choice over the mean-field decomposition since the coefficients $A_j^i$, where $j ~\epsilon~ \{1,2,3,4,5\}, i~\epsilon~\{x,y,z\}$, can take any value. Here we are making one particular choice of decoupling $A_j^i=-\frac{1}{4}$. This ensures that the mean-field energy is bounded from below and that the mean-field Hamiltonian will have all the possible mean-field channels. With the above choice, the Hamiltonian will become,
\begin{equation}
    \begin{split}
    H=&-\frac{3}{4}J\sum_{\langle i,j \rangle}[\{\hat{\chi}_{ij}^\dagger \hat{\chi}_{ij}+\hat{\Delta}_{ij}^\dagger \hat{\Delta}_{ij} +{\hat{E}_{ij}}^{x\dagger} \hat{E}_{ij}^x+{\hat{E}_{ij}}^{y\dagger} \hat{E}_{ij}^y\\
    &+{\hat{E}_{ij}}^{z\dagger} \hat{E}_{ij}^z+\frac{1}{3}({\hat{D}_{ij}}{x^\dagger} \hat{D}_{ij}^x+{\hat{D}_{ij}}^{y\dagger} \hat{D}_{ij}^y+{\hat{D}_{ij}}{z^\dagger} \hat{D}_{ij}^z)\}\\
    &-\frac{3}{4}\hat{n}_i\hat{n}_j+3\hat{n}_j]\\
    \end{split}
\end{equation}
Upon enforcing the half-filling, i.e.
\begin{equation*}
    f_{i\alpha}^\dagger f_{i\alpha}=1
\end{equation*}
the last two terms will become constants and can be dropped, as this will shift the energy spectrum by a constant amount. Which gives, 
\begin{equation}
    \begin{split}
    H_{MF}&=-\frac{3}{4}J\sum_{\langle i,j \rangle}[\{{\chi}_{ij}^* \hat{\chi}_{ij}+{\Delta}_{ij}^* \hat{\Delta}_{ij} +{{E}_{ij}^x}^* \hat{E}_{ij}^x+{{E}_{ij}^y}^* \hat{E}_{ij}^y\\
    &+{{E}_{ij}^z}^* \hat{E}_{ij}^z+\frac{1}{3}({{D}_{ij}^x}^* \hat{D}_{ij}^x+{{D}_{ij}^y}^* \hat{D}_{ij}^y+{{D}_{ij}^z}^* \hat{D}_{ij}^z)+h.c.\}\\
    &-\{\abs{\chi_{ij}}^2+\abs{\Delta_{ij}}^2+\abs{E^x_{ij}}^2+\abs{E^y_{ij}}^2+\abs{E^z_{ij}}^2\\
    &+\frac{1}{3}(\abs{D^x_{ij}}^2+\abs{D^y_{ij}}^2+\abs{D^z_{ij}}^2)\}]\\
    \end{split}
\end{equation}
\section{Symmetry relations between mean-field parameters}
\label{mf symmetric relations}
This section finds the symmetry relations between the mean-field parameters in the Eq. \ref{mf_chi_e}(in the main text), where each bond in the pyrochlore unit cell has four complex mean-field parameters $\hat{\chi}$ and $\hat{E^a}$. Each symmetry operation maps one bond to the other. A detailed list of bond transformation and constraints on mean fields due to the corresponding symmetry operation is given in table \ref{mftable1} and \ref{mftable2}. 
\begin{figure}[ht!]
    \centering
    \includegraphics[width=0.15\textwidth]{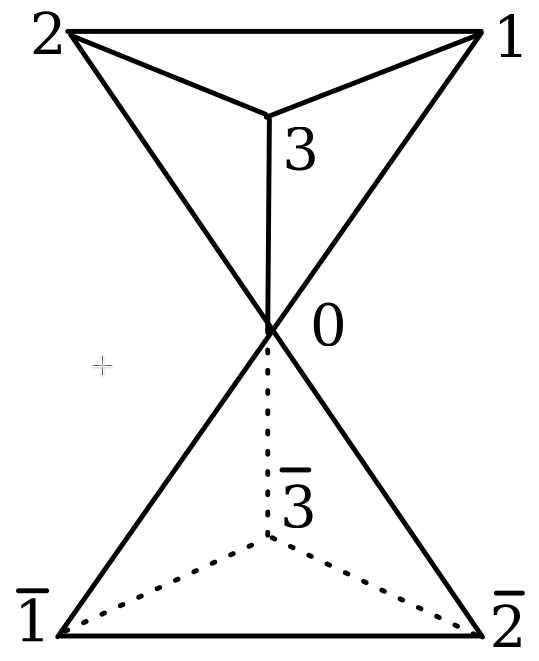}
    \caption{Pyrochlore unit cell with four sublattices indexed with 0, 1, 2, 3 and 12 bonds depticed the figure. The sublattices denoted with $\bar{1}$, $\bar{2}$, and $\bar{3}$, although do not belong to the unit cell, are considered to label the bonds and to show the bond transformation upon symmetry operation.}
\end{figure}

\begin{center}
\begin{table*}[hbt!]
\renewcommand{\arraystretch}{1.5}
    \begin{tabular}{c | c | c | c | c | c | c | c   }
    \hline
    Symmetry   & Bond           &   Case 1 & Case 2 & Case 3 & Case 4&  Case 7 & Case 8 \\
    Operations & Transformation &         &\\
    \hline
    $\mathbf{(23)}$&$(01)\rightarrow(01)$ & $u_{01}^x=0$& $u_{01}^x=0$&$u_{01}^0=0$&$u_{01}^0=0$&$u_{01}^x=0$&$u_{01}^x=0$\\
    &&$u_{01}^z=0$& $u_{01}^z=0$&$u_{01}^y=0$&$u_{01}^y=0$&$u_{01}^z=0$&$u_{01}^z=0$ \\
    \hline
    $\mathbf{(14)}$ & $(01) \rightarrow (10)$ & $u_{01}^y=0$& $u_{01}^y=0$&$u_{01}^{x(z)}=-u_{10}^{x(z)}$&$u_{01}^{x(z)}=-u_{10}^{x(z)}$&$\sigma^1 u_{01}^y\sigma^1=u_{10}^y$&$\sigma^1 u_{01}^y\sigma^1=u_{10}^y$\\    
    \hline
    $\mathbf{(14)(23)}$ & $(01) \rightarrow (10)$ & $u_{01}^\alpha = u_{10}^\alpha$& $u_{01}^y=0$&$u_{01}^\alpha=-u_{10}^\alpha$&$u_{01}^\alpha=-u_{10}^\alpha$&$u_{01}^0=0$&$u_{01}^0=0$\\
    \hline
    $C_3 = \bar{C}_6^{-2}$ & $(0 1) \rightarrow (0 2)$ & $u_{01}^\alpha = u_{02}^\alpha$&$u_{01}^0 = -u_{02}^0$ & $u_{01}^{x(z)} = u_{02}^{x(z)}$&$u_{01}^{x(z)} = -u_{02}^{x(z)}$&$u_{01}^y = u_{02}^y$&$u_{01}^y = -u_{02}^y$\\
         & $(0 \bar{1}) \rightarrow (0 \bar{2})$ & $u_{0\bar{1}}^\alpha = u_{0\bar{2}}^\alpha$&$u_{0\bar{1}}^0 = -u_{0\bar{2}}^0$& $u_{0\bar{1}}^{x(z)} = u_{0\bar{2}}^{x(z)}$&$u_{0\bar{1}}^{x(z)} = -u_{0\bar{2}}^{x(z)}$& $u_{0\bar{1}}^y = u_{0\bar{2}}^y$&$u_{0\bar{1}}^y =- u_{0\bar{2}}^y$\\
         & $(1 2) \rightarrow (2 3)$ & $u_{12}^\alpha = u_{23}^\alpha$&$u_{12}^0 = -u_{23}^0$& $u_{12}^{x(z)} = u_{23}^{x(z)}$&$u_{12}^{x(z)} = -u_{23}^{x(z)}$&$u_{12}^y = u_{23}^y$&$u_{12}^y =- u_{23}^y$\\
         & $(\bar{1} \bar{2}) \rightarrow (\bar{2} \bar{3}) $ &$u_{\bar{1}\bar{2}}^\alpha = u_{\bar{2}\bar{3}}^\alpha$&$u_{\bar{1}\bar{2}}^0 = -u_{\bar{2}\bar{3}}^0$& $u_{\bar{1}\bar{2}}^{x(z)} = u_{\bar{2}\bar{3}}^{x(z)}$&$u_{\bar{1}\bar{2}}^{x(z)} = -u_{\bar{2}\bar{3}}^{x(z)}$&$u_{\bar{1}\bar{2}}^y = u_{\bar{2}\bar{3}}^y$&$u_{\bar{1}\bar{2}}^y = -u_{\bar{2}\bar{3}}^y$\\
    \hline
    $C_3^2=\bar{C}_6^{2}$& $(0 1) \rightarrow (0 3)$& $u_{01}^\alpha = u_{03}^\alpha$& $u_{01}^0 = -u_{03}^0$&$u_{01}^{x(z)} = u_{03}^{x(z)}$& $u_{01}^{x(z)} = -u_{03}^{x(z)}$& $u_{01}^y = u_{03}^y$& $u_{01}^y = -u_{03}^y$\\
     & $(0 \bar{1}) \rightarrow (0 \bar{3})$ & $u_{0\bar{1}}^\alpha = u_{0\bar{3}}^\alpha$&$u_{0\bar{1}}^0 = -u_{0\bar{3}}^0$&$u_{0\bar{1}}^{x(z)} = u_{0\bar{3}}^{x(z)}$&$u_{0\bar{1}}^{x(z)} =-u_{0\bar{3}}^{x(z)}$&$u_{0\bar{1}}^y = u_{0\bar{3}}^y$&$u_{0\bar{1}}^y = -u_{0\bar{3}}^y$\\
    &$(1 2) \rightarrow (31)$ & $u_{12}^\alpha = u_{31}^\alpha$&$u_{12}^0 = u_{31}^0$&$u_{12}^{x(z)} = u_{31}^{x(z)}$&$u_{12}^{x(z)} = u_{31}^{x(z)}$&$u_{12}^y = u_{31}^y$&$u_{12}^y = u_{31}^y$\\
    &$(\bar{1} \bar{2}) \rightarrow (\bar{3} \bar{1}) $ &$u_{\bar{1}\bar{2}}^\alpha = u_{\bar{3}\bar{1}}^\alpha$&$u_{\bar{1}\bar{2}}^0= u_{\bar{3}\bar{1}}^0$&$u_{\bar{1}\bar{2}}^{x(z)} = u_{\bar{3}\bar{1}}^{x(z)}$&$u_{\bar{1}\bar{2}}^{x(z)} = u_{\bar{3}\bar{1}}^{x(z)}$&$u_{\bar{1}\bar{2}}^y = u_{\bar{3}\bar{1}}^y$&$u_{\bar{1}\bar{2}}^y = u_{\bar{3}\bar{1}}^y$\\
    \hline
     $I = \bar{C}_6^{3}$ &  $(0 1) \rightarrow (0 \bar{1})$& $u_{01}^\alpha = u_{0\bar{1}}^\alpha$&$u_{01}^0 = -u_{0\bar{1}}^0$& $u_{01}^{x(z)} = u_{0\bar{1}}^{x(z)}$&$u_{01}^{x(z)} = -u_{0\bar{1}}^{x(z)}$&$u_{01}^y = u_{0\bar{1}}^y$&$u_{01}^y =- u_{0\bar{1}}^y$\\
      & $({1} {2}) \rightarrow (\bar{1} \bar{2}) $& $u_{12}^\alpha = u_{\bar{1}\bar{2}}^\alpha$&$u_{12}^0 = u_{12}^0$&$u_{12}^{x(z)} = u_{\bar{1}\bar{2}}^{x(z)}$&$u_{12}^{x(z)} = u_{\bar{1}\bar{2}}^{x(z)}$&$u_{12}^y = u_{\bar{1}\bar{2}}^y$&$u_{12}^y = u_{\bar{1}\bar{2}}^y$\\
    \hline
    $\Sigma = S.\mathcal{I}$ & $(01)\rightarrow(31)$ & $u_{01}^\alpha = u_{31}^\alpha$&$u_{01}^0 = u_{31}^0$ & $u_{01}^{x(z)} = u_{31}^{x(z)}$&$u_{01}^{x(z)} = u_{31}^{x(z)}$&$u_{31}^y=-\sigma^1u_{01}^y\sigma^1$&$u_{31}^y=-\sigma^1u_{01}^y\sigma^1$ \\
    \hline
    \end{tabular}

\caption{The constraints on mean-field are given for PSG cases 1 to 8. The PSG cases 5 and 6 are not mentioned as all the mean fields are zero.}
\label{mftable1}
\end{table*}
\end{center}

\begin{center}
\begin{table*}[hbt!]

\renewcommand{\arraystretch}{1.5}
    \begin{tabular}{c | c | c | c | c | c | c | c   }
    \hline
    Symmetry   & Bond           &   Case 11 & Case 12 & Case 13 & Case 14&  Case 15 & Case 16 \\
    Operations & Transformation &         &\\
    \hline
    $\mathbf{(23)}$&$(01)\rightarrow(01)$ &$u_{01}^x=0$&$u_{01}^x=0$&$u_{01}^x=0$&$u_{01}^x=0$&$u_{01}^0=0$&$u_{01}^0=0$\\
    &&$u_{01}^z=0$& $u_{01}^z=0$& $u_{01}^z=0$&$u_{01}^z=0$&$u_{01}^y=0$&$u_{01}^y=0$ \\
    \hline
    $\mathbf{(14)}$ & $(01) \rightarrow (10)$ &$\sigma^1 u_{01}^y\sigma^1=u_{10}^y$ & $\sigma^1 u_{01}^y\sigma^1=u_{10}^y$ &$u_{01}^y=0$&$u_{01}^y=0$&$u_{01}^{x(z)}=-u_{10}^{x(z)}$&$u_{01}^{x(z)}=-u_{10}^{x(z)}$\\    
    \hline
    $\mathbf{(14)(23)}$ & $(01) \rightarrow (10)$ &$u_{01}^0=0$ & $u_{01}^0=0$&$u_{01}^\alpha=u_{10}^\alpha$&$u_{01}^\alpha=u_{10}^\alpha$&$u_{01}^\alpha = -u_{10}^\alpha$&$u_{01}^\alpha =-u_{10}^\alpha$\\
    \hline
    $C_3 = \bar{C}_6^{-2}$ & $(0 1) \rightarrow (0 2)$&$u_{01}^y = u_{02}^y$&$u_{01}^y = -u_{02}^y$ & $u_{01}^0 = u_{02}^0$&$u_{01}^0 = -u_{02}^0$ & $u_{01}^{x(z)} = u_{02}^{x(z)}$&$u_{01}^{x(z)} = -u_{02}^{x(z)}$\\
         & $(0 \bar{1}) \rightarrow (0 \bar{2})$& $u_{0\bar{1}}^y = u_{0\bar{2}}^y$&$u_{0\bar{1}}^y =- u_{0\bar{2}}^y$ & $u_{0\bar{1}}^0 = u_{0\bar{2}}^0$&$u_{0\bar{1}}^0 = -u_{0\bar{2}}^0$& $u_{0\bar{1}}^{x(z)} = u_{0\bar{2}}^{x(z)}$&$u_{0\bar{1}}^{x(z)} = -u_{0\bar{2}}^{x(z)}$\\
         & $(1 2) \rightarrow (2 3)$ &$u_{12}^y = u_{23}^y$&$u_{12}^y =- u_{23}^y$& $u_{12}^0 = u_{23}^0$&$u_{12}^0 = -u_{23}^0$& $u_{12}^{x(z)} = u_{23}^{x(z)}$&$u_{12}^{x(z)} = -u_{23}^{x(z)}$\\
         & $(\bar{1} \bar{2}) \rightarrow (\bar{2} \bar{3}) $ &$u_{\bar{1}\bar{2}}^y = u_{\bar{2}\bar{3}}^y$&$u_{\bar{1}\bar{2}}^y = -u_{\bar{2}\bar{3}}^y$&$u_{\bar{1}\bar{2}}^0 = u_{\bar{2}\bar{3}}^0$&$u_{\bar{1}\bar{2}}^0 = -u_{\bar{2}\bar{3}}^0$& $u_{\bar{1}\bar{2}}^{x(z)} = u_{\bar{2}\bar{3}}^{x(z)}$&$u_{\bar{1}\bar{2}}^{x(z)} = -u_{\bar{2}\bar{3}}^{x(z)}$\\
    \hline
    $C_3^2=\bar{C}_6^{2}$& $(0 1) \rightarrow (0 3)$& $u_{01}^y = u_{03}^y$& $u_{01}^y = -u_{03}^y$& $u_{01}^0 = u_{03}^0$& $u_{01}^0 = -u_{03}^0$&$u_{01}^{x(z)} = u_{03}^{x(z)}$& $u_{01}^{x(z)} = -u_{03}^{x(z)}$\\
     & $(0 \bar{1}) \rightarrow (0 \bar{3})$ &$u_{0\bar{1}}^y = u_{0\bar{3}}^y$&$u_{0\bar{1}}^y = -u_{0\bar{3}}^y$ &$u_{0\bar{1}}^0 = u_{0\bar{3}}^0$&$u_{0\bar{1}}^0 = -u_{0\bar{3}}^0$&$u_{0\bar{1}}^{x(z)} = u_{0\bar{3}}^{x(z)}$&$u_{0\bar{1}}^{x(z)} =-u_{0\bar{3}}^{x(z)}$\\
    &$(1 2) \rightarrow (31)$ & $u_{12}^y = u_{31}^y$&$u_{12}^y = u_{31}^y$&$u_{12}^0 = u_{31}^0$&$u_{12}^0 = u_{31}^0$&$u_{12}^{x(z)} = u_{31}^{x(z)}$&$u_{12}^{x(z)} = u_{31}^{x(z)}$\\
    &$(\bar{1} \bar{2}) \rightarrow (\bar{3} \bar{1}) $&$u_{\bar{1}\bar{2}}^y = u_{\bar{3}\bar{1}}^y$&$u_{\bar{1}\bar{2}}^y = u_{\bar{3}\bar{1}}^y$ &$u_{\bar{1}\bar{2}}^0 = u_{\bar{3}\bar{1}}^0$&$u_{\bar{1}\bar{2}}^0= u_{\bar{3}\bar{1}}^0$&$u_{\bar{1}\bar{2}}^{x(z)} = u_{\bar{3}\bar{1}}^{x(z)}$&$u_{\bar{1}\bar{2}}^{x(z)} = u_{\bar{3}\bar{1}}^{x(z)}$\\
    \hline
     $I = \bar{C}_6^{3}$ &  $(0 1) \rightarrow (0 \bar{1})$&$u_{01}^y = -u_{0\bar{1}}^y$&$u_{01}^y = u_{0\bar{1}}^y$& $u_{01}^0 = -u_{0\bar{1}}^0$&$u_{01}^0 = u_{0\bar{1}}^0$& $u_{01}^{x(z)} =- u_{0\bar{1}}^{x(z)}$&$u_{01}^{x(z)} = u_{0\bar{1}}^{x(z)}$\\
      & $({1} {2}) \rightarrow (\bar{1} \bar{2}) $&$u_{12}^y =- u_{\bar{1}\bar{2}}^y$&$u_{12}^y = -u_{\bar{1}\bar{2}}^y$& $u_{12}^0 = -u_{\bar{1}\bar{2}}^0$&$u_{12}^0 = -u_{12}^0$&$u_{12}^{x(z)} = -u_{\bar{1}\bar{2}}^{x(z)}$&$u_{12}^{x(z)} = -u_{\bar{1}\bar{2}}^{x(z)}$\\
    \hline
    $\Sigma = S.\mathcal{I}$ & $(01)\rightarrow(31)$&$u_{31}^y=-\sigma^1u_{01}^y\sigma^1$&$u_{31}^y=-\sigma^1u_{01}^y\sigma^1$ & $u_{01}^0 = u_{31}^0$&$u_{01}^0 = u_{31}^0$ & $u_{01}^{x(z)} =- u_{31}^{x(z)}$&$u_{01}^{x(z)} = u_{31}^{x(z)}$ \\
    \hline
    \end{tabular}

\caption{The constraints on mean-field are given for PSG cases 11 to 16. The PSG cases 9 and 10 are not mentioned as all the mean-field are zero.}
\label{mftable2}
\end{table*}
\end{center}

Given a bond $\langle ij \rangle $, first we consider the Hamiltonian on that bond:
\begin{equation*}
    \text{Tr}[{\sigma^\alpha\Psi_{r_i}u_{r_ir_j}^\alpha\Psi_{r_j}^{\dagger}}].
\end{equation*}
Then, we perform the PSG operation corresponding to the symmetry operation, say 'A', and we get,
\begin{equation*}
    \text{Tr}[{\sigma^\alpha U_A^\dagger \Psi_{A(r_i)}W_A^{\dagger}(A(r_i))u_{r_ir_j}^\alpha W_A(A(r_j)) \Psi_{A(r_j)}^{\dagger}}U_A]
\end{equation*}
We already know how the bonds are going to be mapped under A, i.e., A(i,j)=($i^{\prime},j^{\prime}$)(table \ref{mftable1} and \ref{mftable2}). The above term can be simplified as,
\begin{equation*}
    \text{Tr}[{\sigma^\alpha U_A^\dagger \Psi_{r_i^{\prime}}W_A^{\dagger}(r_i^{\prime})u_{r_ir_j}^\alpha W_A(r_j^{\prime}) \Psi_{r_j^{\prime}}^{\dagger}}U_A]
\end{equation*}
We can compare this to the Hamiltonian on the bond $\langle ij \rangle$ as,
\begin{equation*}
    \text{Tr}[{\sigma^\alpha\Psi_{r_i^{\prime}}u_{r_i^{\prime}r_j^{\prime}}^\alpha \Psi_{r_j^{\prime}}^{\dagger}}]= \text{Tr}[{\sigma^\alpha U_A^\dagger \Psi_{r_i^{\prime}}W_A^{\dagger}(r_i^{\prime})u_{r_ir_j}^\alpha W_A(r_j^{\prime}) \Psi_{r_j^{\prime}}^{\dagger}}U_A]
\end{equation*}
which establishes the relation between,$u_{r_ir_j}$ and $u_{r_i^{\prime}r_j^{\prime}}$.

\begin{figure}[ht!]
    \centering
    \includegraphics[width=0.5\textwidth]{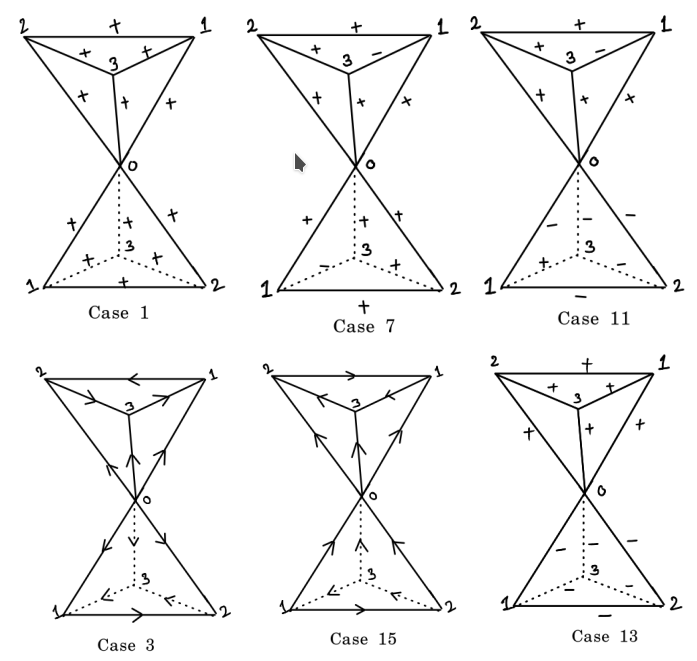}
    \caption{The PSG allowed mean-fields are denoted as the +/- sign and the arrows, are given regarding the mean-field on $(01)$ bond, over the bonds of the pyrochlore unit cell. +/- sign indicates the phase 0/$\pi$ picked up by the spinon upon going through the bond. Case 3 and 15 are monopole flux states, where the arrows indicate that the spinon picks up a phase of $\frac{\pi}{2}$ when it hops along the arrow and a phase of $-\frac{\pi}{2}$ when it moves opposite to the arrow.} 
    \label{odd_mf_figure}
\end{figure}

We find these relations for each symmetry operation listed in table \ref{mftable1} and \ref{mftable2}, 
on the bonds for each of the 16 PSG classes. This will put forth the constraints on the mean-field parameters, which are given in table \ref{mftable1} and \ref{mftable2} and also is decorated as the +/- signs and arrows over the bonds of the unit cell of pyrochlore in Fig(\ref{odd_mf_figure}).  Moreover, we can use the translation PSG to establish the relation between mean-field in different unit cells. Note that when $\chi_1=0$, the mean-fields are equal in every unit cell, i.e.,
\begin{equation*}
    u^{\alpha}_{T_i(\mathbf{r}_\mu,{\mathbf{r}^\prime_\nu})} = u^{\alpha}_{r_\mu,r_\nu},~~~i=1,2,3
\end{equation*}
However, for $\chi_1=\pi$, 
\begin{equation*}
    \begin{split}
        & T_1:u^{\alpha}_{\mathbf{r}_\mu,{\mathbf{r}^\prime}_\nu}=u^{\alpha}_{T_1(\mathbf{r}_\mu),T_1({\mathbf{r}^\prime}_\nu)} \\
        & T_2:u^{\alpha}_{T_2(\mathbf{r}_\mu),{T_2(\mathbf{r}^\prime}_\nu)}=(-1)^{r_1+r_1^{\prime}}u^{\alpha}_{\mathbf{r}_\mu,{\mathbf{r}^\prime}_\nu} \\
        & T_3:u^{\alpha}_{T_3(\mathbf{r}_\mu),{T_3(\mathbf{r}^\prime}_\nu)}=(-1)^{r_1+r_1^{\prime}}(-1)^{r_2+r_2^{\prime}}u^{\alpha}_{\mathbf{r}_\mu,{\mathbf{r}^\prime}_\nu} \\
    \end{split}
\end{equation*}
Since, $T_1(r_1,r_2,r_3)_{\mu}=(r_1+1,r_2,r_3)_{\mu}, T_2(r_1,r_2,r_3)_{\mu}=(r_1,r_2+1,r_3)_{\mu}, T_3(r_1,r_2,r_3)_{\mu}=(r_1,r_2,r_3+1)_{\mu}$, doing a translation along $T_2$ does not change $r_1, r_3$. Hence doing two consecutive translations along $T_2$ restores $u^{\alpha}_{\mathbf{r}_\mu,{\mathbf{r}^\prime}_\nu}$. Similarly, doing a $T_3$ operation does not affect $r_1, r_2$ and resulting in restoring the $u^{\alpha}_{\mathbf{r}_\mu,{\mathbf{r}^\prime}_\nu}$ upon two successive $T_3$ operation.
The extended pyrochlore unit cell consists of four down-pointing tetrahedra, denoted as bottom, left, right, and top. There are 16 sub-lattices, four from each down-pointing tetrahedra, labeled as B0, B1, B2, B3, L0, L1, L2, L3, R0, R1, R2, R3, T0, T1, T2, T3.
In PSG cases 1, 2, 13, and 14 the mean-field Hamiltonian is
\begin{equation*}
 H_{MF}=-\frac{3}{4}J\left(\sum_{\langle \mathbf{r}_{\mu},\mathbf{r}_{\nu}^{\prime}\rangle}{\chi}_{\mathbf{r}_{\mu},\mathbf{r}_{\nu}^{\prime}}\hat{\chi}_{\mathbf{r}_{\mu},\mathbf{r}_{\nu}^{\prime}}+h.c.\right)+\frac{9}{4}JN\chi^2 
\end{equation*}
Similarly, for PSG cases 3, 4, 15 and 16 
\begin{eqnarray*}
    H_{MF}&=&-\frac{3}{4}J\left(\sum_{\langle i,j \rangle}{{E}_{ij}^x}^* \hat{E}_{ij}^x
    +{{E}_{ij}^z}^* \hat{E}_{ij}^z+\text{h.c.} \right) \\
    &+& \frac{3}{4}JN(\abs{E^x}^2+\abs{E^z}^2)]  
\end{eqnarray*}

For PSG cases 7, 8, 11, and 12 the mean-field Hamiltonian will be,
\begin{equation*}
    H_{MF}=-\frac{3}{4}J\left(\sum_{\langle i,j \rangle}{{E}_{ij}^y}^* \hat{E}_{ij}^y
    +\text{h.c.}\right)+ \frac{9}{4}JN\abs{E^y}^2
\end{equation*}
\begin{figure*}[!hbt]     
  \includegraphics[width=0.3\textwidth]{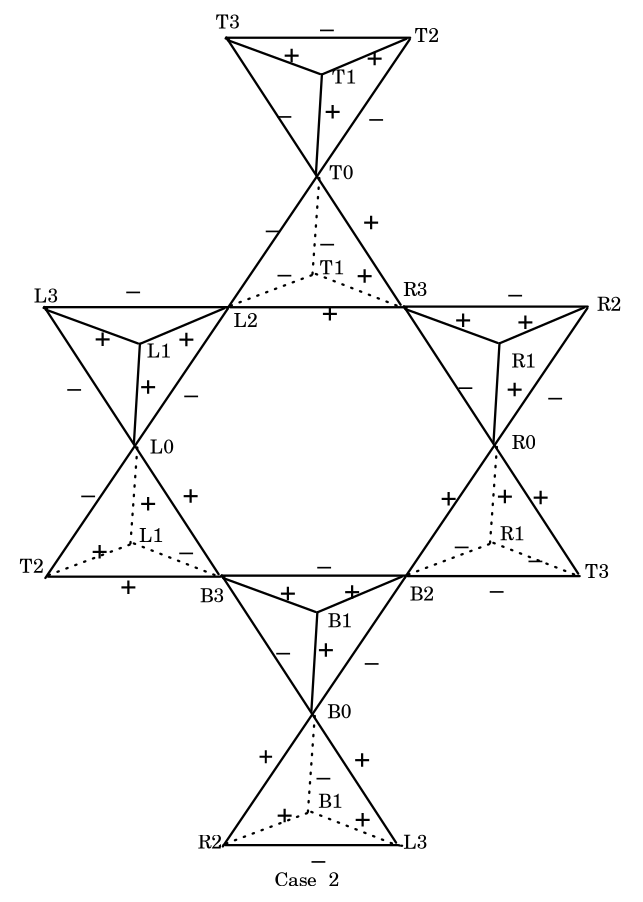}
    \hspace{0.1cm} 
     \includegraphics[width=0.3\textwidth]{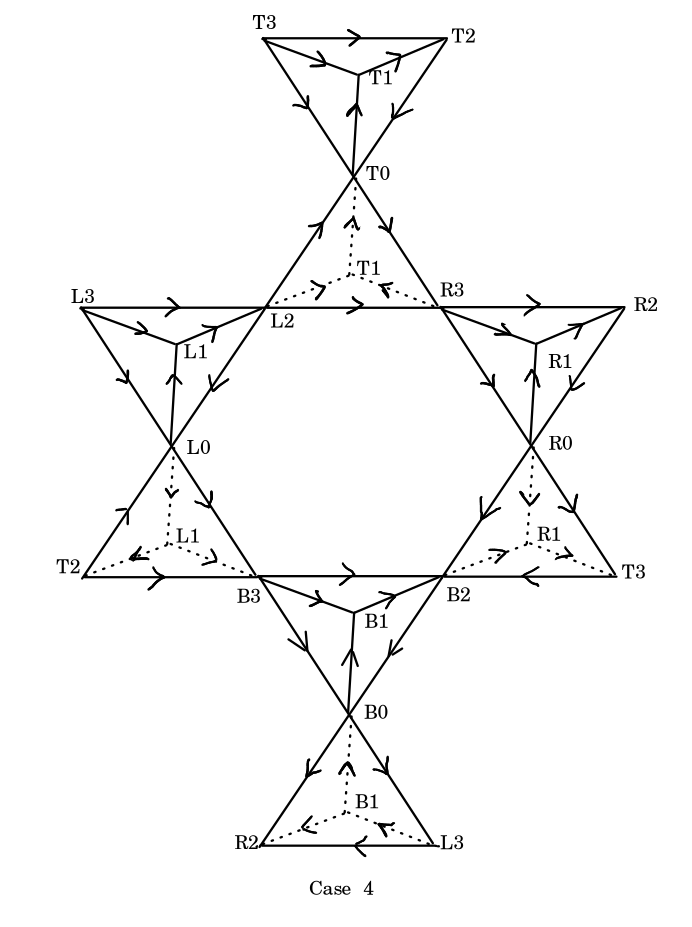}
     \hspace{0.1cm}
     \includegraphics[width=0.3\textwidth]{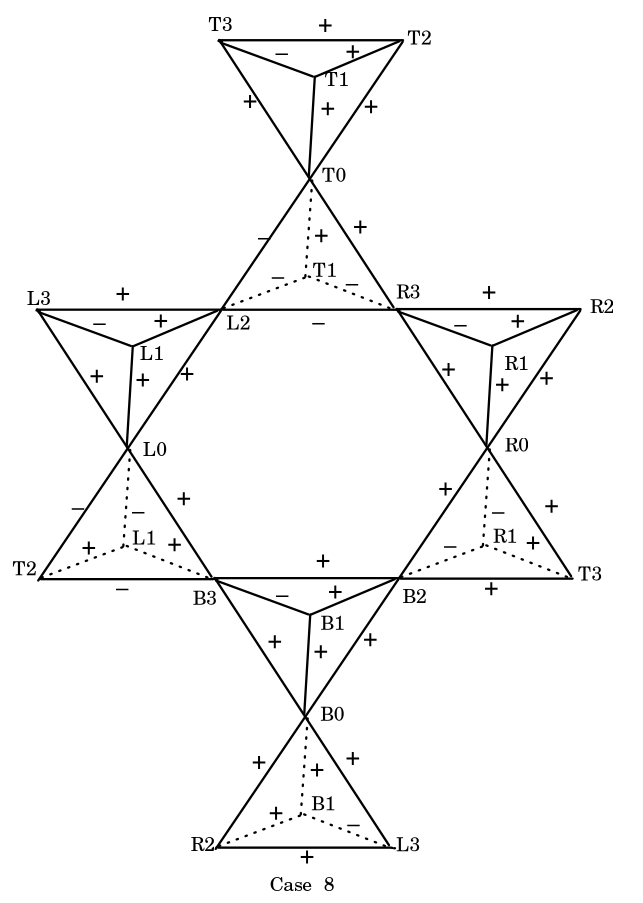}
     \includegraphics[width=0.3\textwidth]{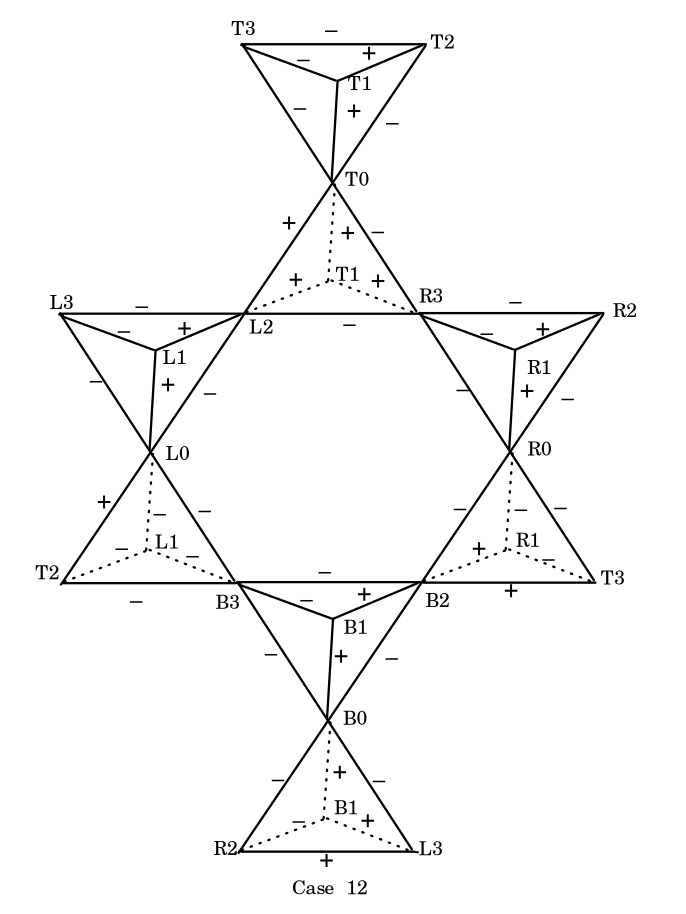}
    \hspace{0.1cm} 
     \includegraphics[width=0.3\textwidth]{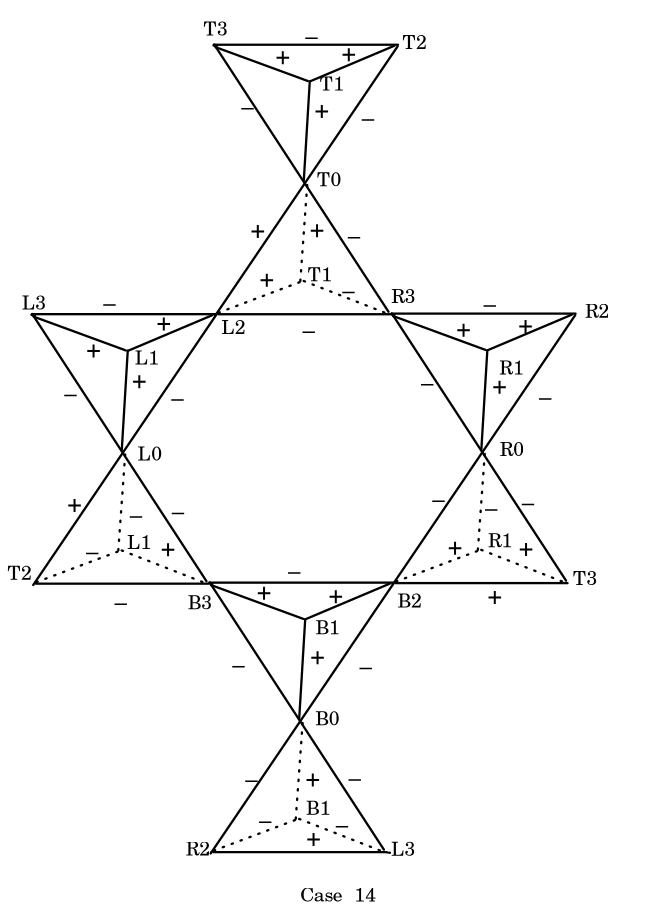}
     \hspace{0.1cm}
     \includegraphics[width=0.3\textwidth]{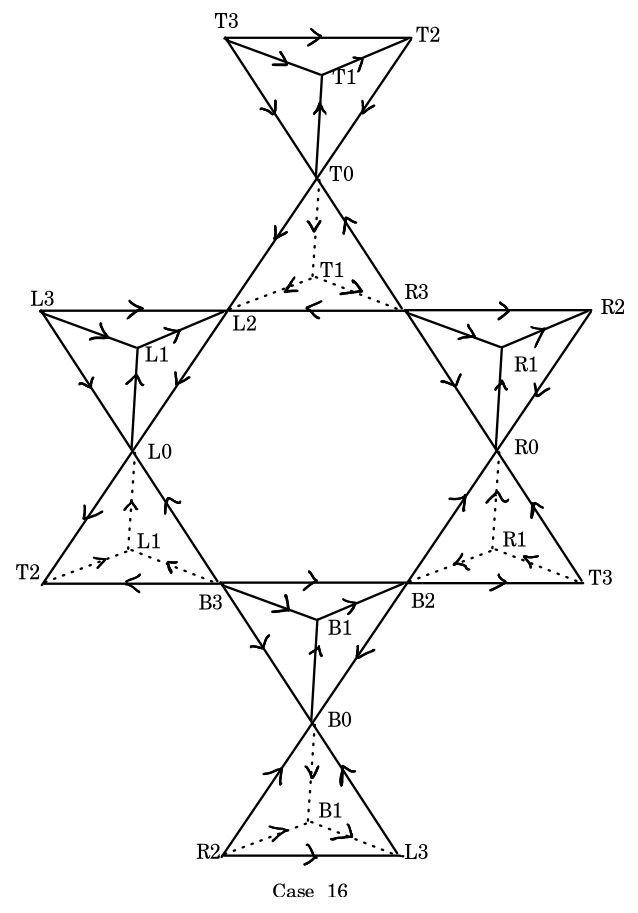}
     \caption{In these figures, we represent the PSG allowed mean-fields for the even cases with +/- signs and the arrows over the bonds, regarding the mean-field on (01) bond.}
     \label{even_mf_figure}  
\end{figure*}

\newpage
\bibliography{full_list_ref}
\end{document}